\documentclass[twocolumn]{aastex631}

\shorttitle{Lyman Continuum Galaxies at $z\sim3.1$}
\shortauthors{Liu et al.}

\graphicspath{{./}}

\usepackage{threeparttable}
\usepackage{booktabs}

\begin{document}

\title{Lyman Continuum Emission from Spectroscopically Confirmed Ly$\alpha$ Emitters at $z\sim3.1$}

\author{Yuchen Liu}
\affiliation{Department of Astronomy, School of Physics, Peking University, Beijing 100871, China}
\affiliation{Kavli Institute for Astronomy and Astrophysics, Peking University, Beijing 100871, China}

\author[0000-0003-4176-6486]{Linhua Jiang}
\affiliation{Department of Astronomy, School of Physics, Peking University, Beijing 100871, China}
\affiliation{Kavli Institute for Astronomy and Astrophysics, Peking University, Beijing 100871, China}

\author{Rogier A. Windhorst}
\affiliation{School of Earth and Space Exploration, Arizona State University, Tempe, AZ 85287, USA}

\author{Yucheng Guo}
\affiliation{Department of Astronomy, School of Physics, Peking University, Beijing 100871, China}
\affiliation{Univ Lyon, Univ Lyon1, Ens de Lyon, CNRS, Centre de Recherche Astrophysique de Lyon UMR5574, F-69230, Saint-Genis-Laval, France}

\author[0000-0002-9634-2923]{Zhen-Ya Zheng}
\affiliation{Key Laboratory for Research in Galaxies and Cosmology, Shanghai Astronomical Observatory, Chinese Academy of Sciences, 80 Nandan Road, Shanghai 200030, People’s Republic of China}

\begin{abstract}

We present a study of Lyman continuum (LyC) emission in a sample of $\sim$150 Ly$\alpha$ emitters (LAEs) at $z\approx3.1$ in the Subaru-XMM Deep Survey field. These LAEs were previously selected using the narrowband technique and  spectroscopically confirmed with Ly$\alpha$ equivalent widths (EWs) $\ge45$ \AA. We obtain deep UV images using a custom intermediate-band filter $U_{\rm J}$ that covers a wavelength range of $3330 \sim 3650$ \AA, corresponding to 810$\sim$890 \AA\ in the rest frame. We detect  5 individual LyC galaxy candidates in the $U_{\rm J}$ band, and their escape fractions ($f_{\rm esc}$) of LyC photons are roughly between 40\% and 80\%. This supports a previous finding that a small fraction of galaxies may have very high $f_{\rm esc}$. 
We find that the $f_{\rm esc}$ values of the  5 LyC galaxies are not apparently correlated with other galaxy properties such as Ly$\alpha$ luminosity and EW, UV luminosity and slope, and star-formation rate (SFR). This is partly due to the fact that these galaxies only represent a small fraction ($\sim3$\%) of our LAE sample. For the remaining LAEs that are not detected in $U_{\rm J}$, we stack their $U_{\rm J}$-band images and constrain their average $f_{\rm esc}$. The upper limit of the average $f_{\rm esc}$ value is about 16\%, consistent with the results in the literature. 
Compared with the non-LyC LAEs, the LyC LAEs tend to have higher Ly$\alpha$ luminosities, Ly$\alpha$ EWs, and SFRs, but their UV continuum slopes are similar to those of other galaxies. 

\end{abstract}

\keywords{High-redshift galaxies (734); Lyman-alpha galaxies (978); Star formation (1569)}

\section{Introduction} \label{sec:intro}

Lyman continuum (LyC) photons with wavelength $\rm \lambda<912$ \r{A} ionize neutral hydrogen in both galactic and extra-galactic environments. They are considered to originate from two main sources, massive stars in star-forming galaxies and active super-massive black holes (SMBHs) in galaxy centers \citep{vacca1996lyman,matthee2016production,smith2020lyman}. In recent years, studies of the early universe reveal that the intergalactic medium (IGM) began a transition from a neutral state to an ionized state at $ z\gtrsim6$ \citep{fan2006constraining, becker2015evidence,mason2018universe,keating2020constraining,bosman2022hydrogen}, and sufficient LyC photons are required to complete this reionization process. As the spatial density of AGN/quasars declines rapidly towards high redshift, AGN cannot contribute enough ionizing photons to the UV background at $ z\gtrsim6$ \citep{hopkins2007observational,parsa2018no,faisst2022joint}. \cite{jiang2022definitive} has confirmed from observational data that the AGN population provided a negligible fraction of the total photons required for reionization, and suggested that low-luminosity star-forming galaxies are the dominant ionizing sources.

The contribution of star-forming galaxies critically depends on the fraction ($f_{\rm esc}$) of LyC photons escaping into the IGM. It was suggested that $f_{\rm esc}\rm \mbox{$\sim$} 10\mbox{-}20 \%$ is needed to keep the IGM ionized during the reionization epoch \citep{finkelstein2012candels,bouwens2016lyman,naidu2020rapid}. The required $f_{\rm esc}$ values are also tightly connected with the LyC photon production efficiency in these galaxies. Simulation results have indicated that faint and low-mass galaxies have higher efficiency than brighter and more massive ones \citep{yung2020semi}. Nevertheless, the specific LyC contribution of star-forming galaxies remains unclear. 

Current searches of LyC emitting galaxies are mainly carried out in two redshift windows. Observations of low-redshift (e.g., $z\sim0.3-0.4$) LyC galaxies are made by space telescopes (mainly Hubble Space Telescope, or HST), because they are unfeasible by ground-based telescopes. Several direct detections of low-redshift LyC galaxies show that $f_{\rm esc}$ is around 6\mbox{-}20\%. For example, \cite{izotov2016eight} reported a nearby low-mass galaxy with $f_{\rm esc} \sim 8$\%. Meanwhile, they detected four more LyC galaxies with $f_{\rm esc}=6\mbox{-}13\%$ \citep{izotov2016detection}. \cite{flury2022low} presented a LyC survey of galaxies at $z=0.2\mbox{-}0.4$ that detected 35 LyC emitters with high confidence, and 12 of them have $f_{\rm esc} \ge 5$ \%. These studies usually targeted specific galaxies with strong [O III] line emission that are likely  LyC galaxies.

LyC photons from galaxies at $z\ge3$ can be detected by ground-based observations. \cite{steidel2001lyman} reported the first direct detection of LyC emission from $z\ge3$ galaxies in a composite spectrum. They stacked the spectra of 29 Lyman break galaxies (LBGs) at $z\simeq3.4$ and constrained the galaxy contribution to the radiation field. \cite{shapley2006direct} detected direct ionizing radiation from two individual galaxies at $z\sim3$. So far, dozens of LyC galaxies have been found using direct, spectroscopic observations in the past ten years \citep[e.g.,][]{mostardi2013narrowband,de2016extreme,vanzella2018direct,marques2021uv}. One of the largest samples was provided by the Keck Lyman Continuum Spectroscopic Survey, which detected 15 LyC galaxies at $z\sim3$ and found that their average $f_{\rm esc}$ value was about 9\% \citep{steidel2018keck,pahl2021uncontaminated}.

UV images covering LyC emission provide an alternative method to find LyC galaxies and determine their $f_{\rm esc}$. Deep UV imaging observations have obtained a number of LyC galaxies or candidates \citep{vanzella2012detection,vanzella2016hubble,yuan2021cdfs,saxena2022no}. These galaxies are often Lyman Break Galaxies (LBGs) or LAEs with strong resonant lines like Ly$\rm \alpha$. For galaxies that do not show LyC emission in imaging data, image stacking is an efficient approach to derive their average LyC emission, or constrain the upper limit of their LyC emission \citep{micheva2016searching, grazian2017lyman,begley2022vandels}. A reasonable result from previous studies is $f_{\rm esc}\lesssim 10$\%. 

Due to the increasing IGM opacity, it is very challenging to find LyC galaxies at redshift significantly higher than 3. IGM absorption would result in only a 20\% likelihood of detecting LyC leakage from galaxies at $z\sim4$, and a lower probability  at higher redshifts \citep{inoue2008monte}. Therefore, LyC galaxies at $z\sim 3$ provide an excellent platform to study the properties of ionizing photons and LyC analogs in the early universe.

Previous studies have shown that $f_{\rm esc}$ is correlated with galaxies properties, and some properties can be used as indirect indicators of LyC leakages. For example, LyC galaxies tend to have high star-formation rates, strong resonant lines, and low metallicities. These properties are similar at both low and high redshifts. Ly$\alpha$ emission is potentially a good indicator, as this resonant line emission implies a low H I density state that can allow LyC photons to escape \citep{verhamme2015using,dijkstra2016lyalpha}. There is a positive correlation between $f_{\rm esc}$ and the Ly$\alpha$ emission line equivalent width $W \rm (Ly\alpha)$ \citep{steidel2018keck,pahl2021uncontaminated}. In addition, \cite{furtak2022double} found that the blue-peak of Ly$\alpha$ line can be used as a proxy to the LyC emission. A few other emission lines were also reported to be possible indicators of LyC emitters. For example, the [OIII]/[OII] line ratio is likely related with the leakage of LyC photons \citep{jaskot2019new}. Mg II is a resonant line and is likely correlated with with LyC emission at low redshift \citep{henry2018close,xu2022tracing}. Furthermore, some LyC galaxies tend to have strong C IV emission lines and very weak S II lines \citep{wang2021low,saxena2022strong}. It is worth mentioning that some of the above results are not solid yet due to their small sample sizes and large uncertainties.

In this paper, we present a study of LyC emission from $z\approx 3.1$ LAEs in the Subaru XMM-Newton Deep Survey (SXDS; Figure \ref{fig:coverage}) field, based on our deep UV imaging observations. The LAE sample consists of $\sim$200 narrowband selected, spectroscopically confirmed galaxies. The UV filter is a custom filter that covers a rest-frame $810\sim890$ \AA\ for $z=3.1$ galaxies. The structure of the paper is as follows. In Section 2, we present our observations and data reduction. In Section 3, we show the main results on individual LyC detections and stacking analyses. We discuss our results in Section 4 and summarize the paper in Section 5. Throughout the paper we use a $\Lambda$-dominated flat cosmology with $H_0\rm =70\ km \ s^{-1} \ Mpc^{-1}$, $\rm \Omega _{m}=0.3$, and $\Omega_{\Lambda}=0.7$. All magnitudes are in the AB system.

\section{Observations and data} 

\subsection{Observations and Data Reduction}

\begin{figure}[tb]
\centering
\includegraphics[width=0.5\textwidth]{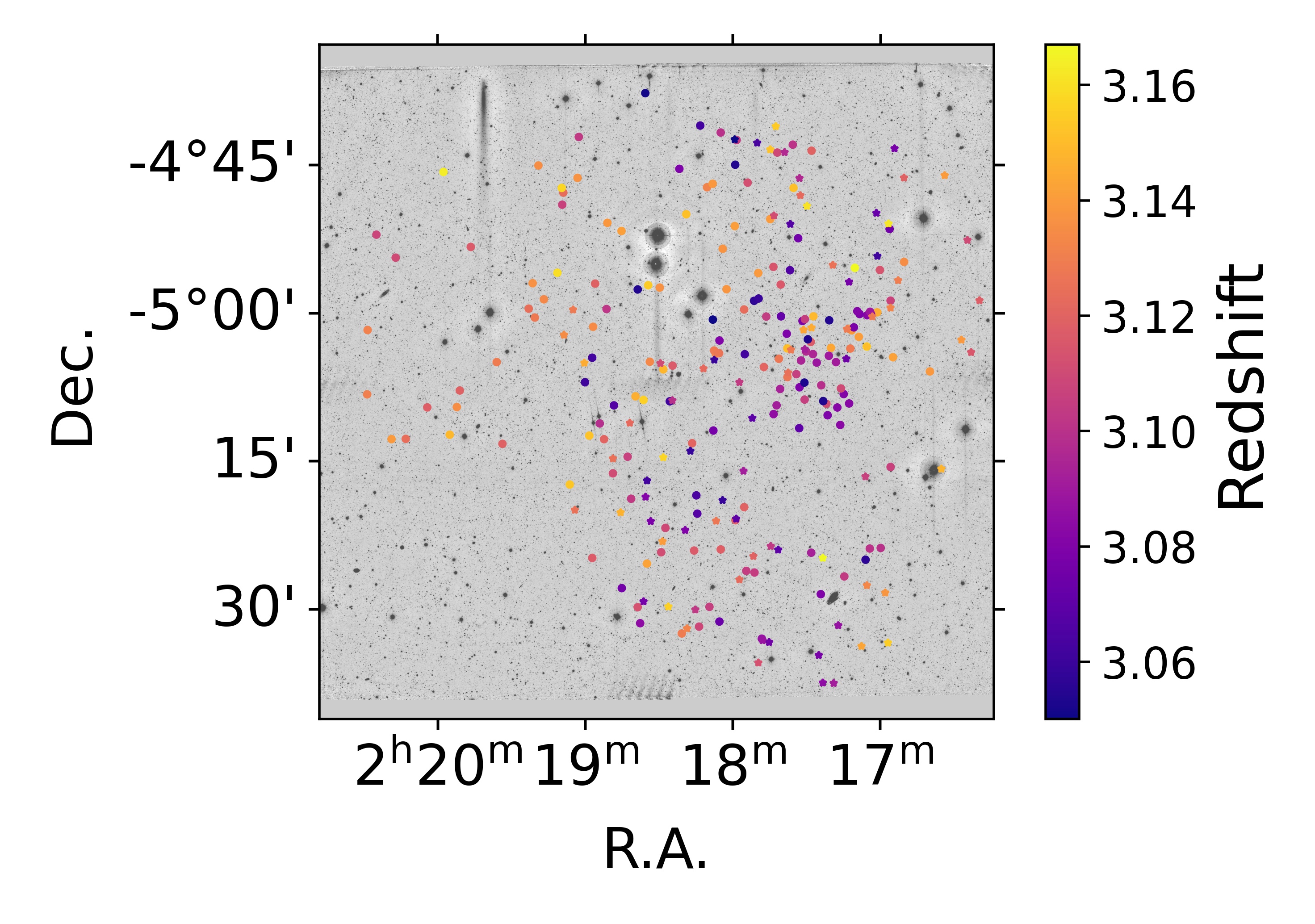}
\caption{The combined $U_{\rm J}$-band image in the SXDS field. The dots represent spectroscopically confirmed LAEs at $z \approx 3.1$. The color bar indicates redshifts. \label{fig:coverage}}
\end{figure}

\begin{figure*}[tb]
\centering
\includegraphics[width=1\textwidth]{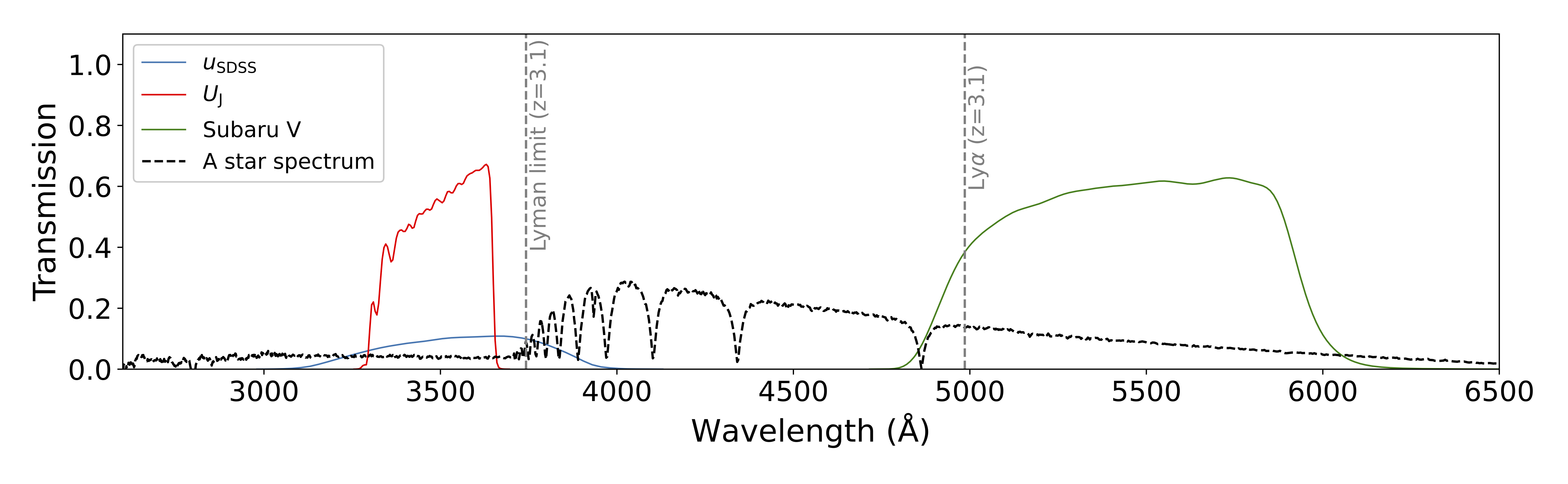}
\caption{Filter transmission curves and a typical spectrum for type A stars. The red curve shows the transmission curve of the $U_{\rm J}$ filter, which is compared with the SDSS $u$ filter in blue. The Subaru $V$ filter is used to estimate the non-ionizing flux. The transmission curves have included the CCD response curves. The dashed line represents a typical A5 star spectrum. The vertical lines indicate the Lyman limit and Ly$\alpha$ emission at $z=3.1$, respectively. \label{fig:exampleA}}
\end{figure*}

The field that we observed is SXDS. This field has deep optical images in a series of broad and narrow bands taken by Subaru Suprime-Cam \citep{furusawa2008subaru} and now by Subaru Hyper-Suprime-Cam \citep{aihara2022third}. The Suprime-Cam imaging data have been widely used to search for LAEs and LBGs at $2<z<7$ \citep{ouchi2008subaru, ouchi2010statistics, curtis2012remarkably, konno2014accelerated, matthee2015identification, jiang2018giant, 2020ApJ...903....4N,guo2020spectroscopic}. 
Our combined images of the Suprime-Cam data reach $27.5\sim28.0$ mag in the $BVRi'$ bands and $>26$ mag in the $z'$ band \citep{jiang2018giant}. \cite{guo2020spectroscopic} selected a large sample of LAE candidates at $z\approx3.1$ based on the images in the $B$, $V$, NB497, and NB503 bands, and built a sample of 179 spectroscopically confirmed LAEs. They also built a robust Ly$\alpha$ luminosity function at $z\sim3.1$, which is consistent with previous studies at the similar redshift range \citep{2012ApJ...744..110C,10.1093/mnras/sty378}. Recently, we obtained 87 more LAEs with spectroscopic redshifts at $z\approx3.1$ in this field. The total number of LAEs at this redshift amounts to 266 and the redshift interval is $3.05\le z \le 3.16$ (Figure \ref{fig:coverage}). The LAE color selection criterion corresponds to a rest-frame equivalent width $\rm EW_{0} \ge 45$\AA. AGNs were removed using the optical spectral data and deep X-ray observations (see details in \cite{guo2020spectroscopic}).
We used a custom filter $U_{\rm J}$ to detect LyC radiation from $z\approx3.1$ LAEs. This filter has a wavelength coverage of $3330 \sim 3650$ \AA\ (wavelengths at half maximum; Figure \ref{fig:exampleA}) with a full width at half maximum (FWHM) of $\sim 320$ \AA. It corresponds to a rest-frame wavelength coverage of 810$\sim$890 \AA\ for our LAEs. This range is very close to the Lyman limit and optimizes the detection of LyC emission, because detectable LyC emission decreases rapidly towards short wavelengths due to IGM absorption. The filter was specially designed to have a sharp cutoff at the red end so that the transmission at $\lambda\approx3700$ \AA\ is nearly zero. This ensures that it does not cover radiation at $\lambda >$ Lyman limit. We further evaluate non-ionizing radiation that leaks into this filter (referred to as ``red-leak'') following \cite{smith2018hubble}. We use the SED model spectrum of the galaxies in our sample to quantify the flux from LyC to UV continuum (see SED fitting in Section 3.3). The result reveals that the maximum contribution of non-ionizing photons to the filter is only $\sim 0.15\%$ for our sample at $z\sim3.1$, so the effect of the ``red-leak" here is negligible.
The $U_{\rm J}$-band observations of the SXDS field were carried out by the 90Prime on the 2.3 m Bok telescope in 2014 and 2015. The 90Prime is a wide-field optical imager with a field-of-view of $1\degr \times 1\degr$. It was equipped with thin CCDs that were optimized for the UV band (at the time of our observations), so it had a high quantum efficiency at the wavelength range that we probed. The observations were made in dark nights/hours with clear skies. The typical seeing, measured by the point spread function (PSF) in $U_{\rm J}$, was $1\farcs5 \sim 2\farcs0$. The typical integration time for individual exposures was 10 min, which ensured that the background noise in the images was dominated by sky background. The total integration time of useful images was 45 hours.

The 90Prime images were reduced in a standard fashion using our own IDL routines. For images taken on each night, we first made a master bias and a master flat image from bias and flat images taken on the same night. We then constructed a bad-pixel mask from the flat image. Bad pixels, saturated pixels, and bleeding trails were later incorporated into weight images. Science images were  corrected for overscan and bias, and were flat-fielded and sky-subtracted. We used SCAMP \citep{bertin2006automatic} to calculate astrometric solutions and used SWARP \citep{bertin2002terapix} to resample and combine all images. The final coadded science image is a weighted average of individual science images, with  a native pixel size of $0\farcs455$. More details can be found in \cite{jiang2015discovery}.

\subsection{Photometry} 

In order to measure the photometry of the LAEs in the $U_{\rm J}$ band, we first derived the magnitude zero point of the $U_{\rm J}$ image. It is not straightforward, since $U_{\rm J}$ is a custom filter. We estimated the zero point using SDSS type A5 stars detected in the SXDS field. Figure \ref{fig:exampleA} shows the spectrum of a standard type A5 star. It is quite flat in the wavelength range that $U_{\rm J}$ covers \citep{gray2002spectroscopic, allende2016new}, and is thus very suitable for our purpose. Figure \ref{fig:exampleA} also shows that the effective wavelengths of $U_{\rm J}$ and the SDSS $u$ (hereafter referred to as $u$) are similar. Our procedure consists of two steps. In the first step, we calculated the difference between $U_{\rm J}$ and $u$ for type A5 stars. We collected HST STIS spectra for a sample of A5 stars \citep{allende2016new} and found that the average value of their $u-U_{\rm J}$ colors was --0.18 mag.

\begin{figure}[tb]
\centering
\includegraphics[width=0.5\textwidth]{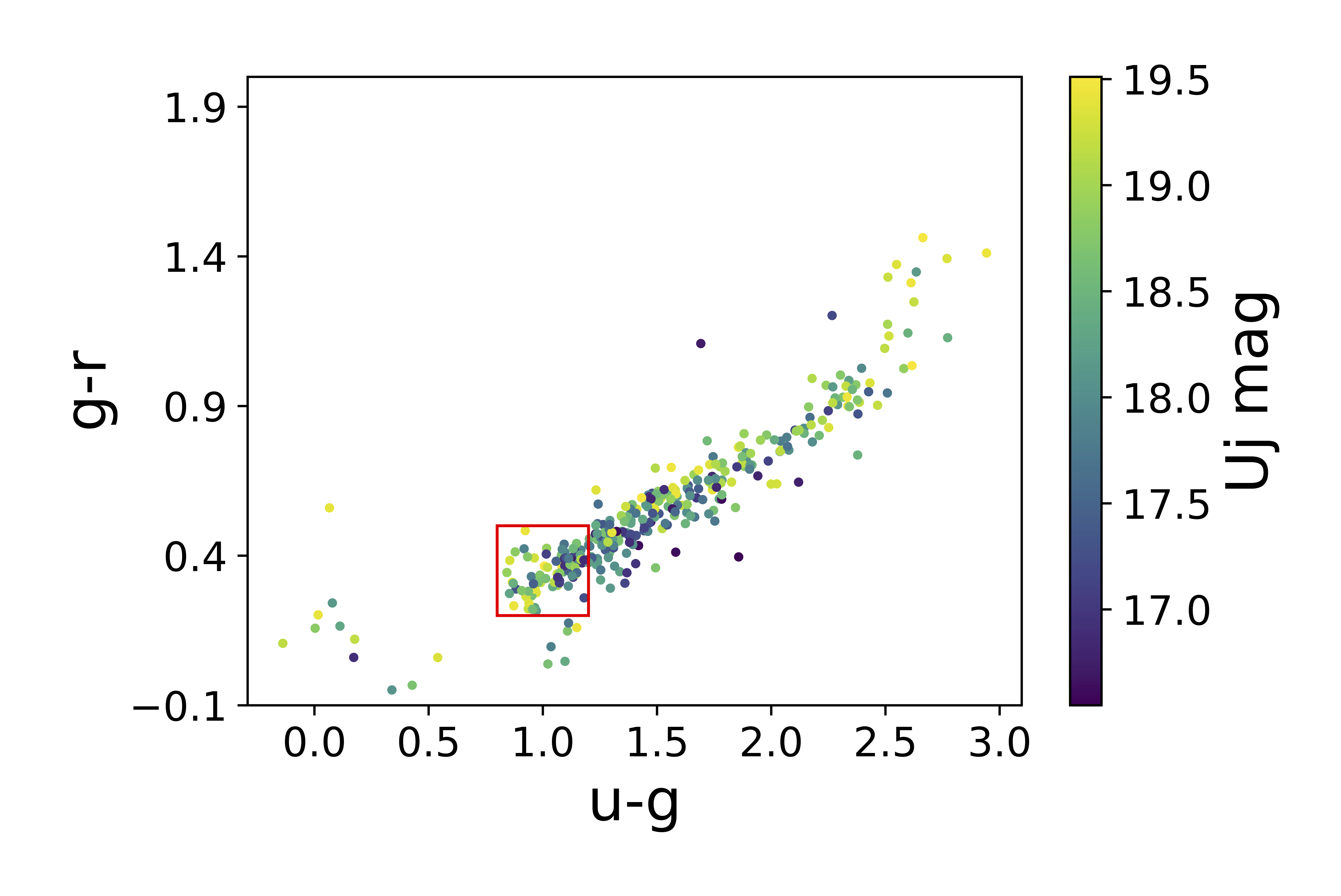}
\caption{The SDSS color-color diagram of bright point sources in the SXDS field. The red rectangle is the selection area for type A stars. The color bar indicates the $U_{\rm J}$ magnitude. \label{fig:2color}}
\end{figure}

In the second step, we used the SDSS $u-g$ vs. $g-r$ color-color diagram to select type A5 stars or stars with a type close to A5. These stars have nearly the same $u-U_{\rm J}$ color. Figure \ref{fig:2color} shows our selection criteria, $0.2$\textless $g-r$\textless $0.5$ and $0.8$\textless $u-g$\textless $1.2$ \citep{covey2007stellar}. The selected objects are bright and point sources. The $U_{\rm J}$-band zero point was determined so that the average $u-U_{\rm J}$ color of these objects was --0.18 mag.

We performed $U_{\rm J}$-band photometry of the LAEs using Python package {\tt photutil}. The majority of the LAEs were not expected to be clearly detected in $U_{\rm J}$, so we performed forced aperture photometry based on the object positions from the narrowband images. The aperture size was $3\farcs6$ in diameter, roughly twice the PSF FWHM. Aperture corrections were obtained from bright point sources and applied to the photometry. The $3\sigma$ detection limit is about 26.8 mag. Small spatial offsets among LyC emission, UV/optical continuum emission, and narrowband emission have been found in previous studies \citep{micheva2016searching}. Our aperture size is large enough so that these offsets are generally negligible.

\section{Results}

\begin{figure}[tb]
\centering
\includegraphics[width=0.45\textwidth]{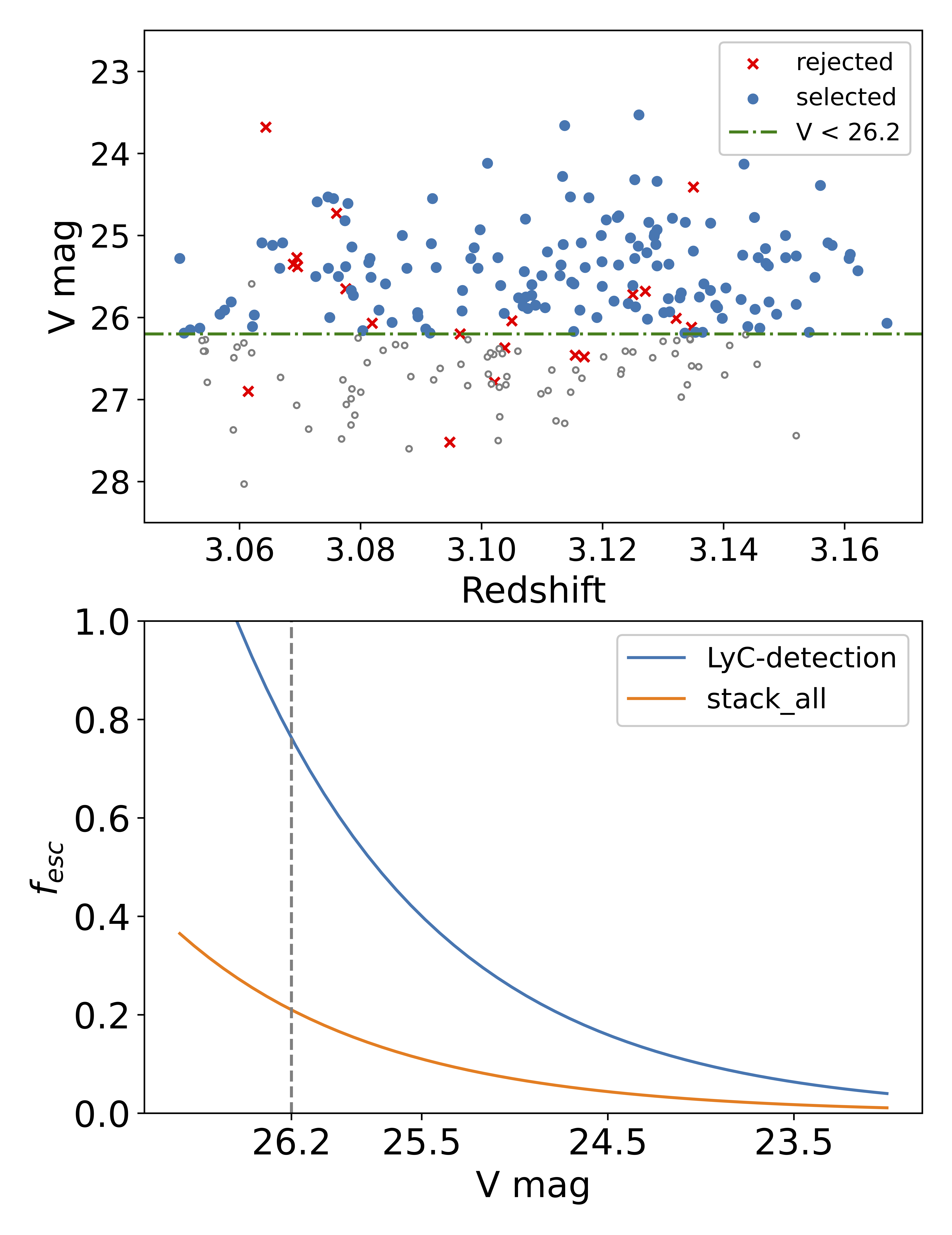}
\caption{LAEs used in this work. The upper panel shows the $V$-band magnitudes and redshifts of the LAEs. The symbols indicate all LAEs in SXDS. The dashed line is our selection criterion of $V=26.2$ mag. Galaxies brighter than $V=26.2$ mag (in blue) are used in this work. The red crosses represent the galaxies that are rejected due to the contamination from their neighbors in the $U_{\rm J}$ band. The lower panel shows the detection limits of $f_{\rm esc}$ for the individual LyC galaxies and stacked galaxies assuming an average IGM absorption and an average dust attenuation (see details in section 3). \label{fig:magz}}
\end{figure}

\begin{deluxetable*}{ccccccccccccc}
\tablecaption{Results of the LyC Measurements}
\tablewidth{0pt}
\setlength{\tabcolsep}{0.8mm}
\label{result}
 \tablehead{
 ID & Redshift&  $U_{\rm J}$ & $\rm M_{UV}$ & L(Ly$\alpha)$ & EW & $\beta$ & $L_{900}/L_{1400}$ &SFR  & $E(B-V)$  & $f_{\rm abs,m}$ & $f_{\rm abs,c}$ \\
  & & & & ($10^{42}$ erg/s)& & & $(M_{\odot}$/yr)& & & & }
\startdata
 LAE1 & 3.167    & 26.53 $\pm$  0.27 &--20.94 & 5.30   &125.10  & $\ldots$ & $\ldots$ & $\ldots$ & $\ldots$ &  $\ldots$& 0.80 $\pm$ 0.26 \\
 LAE2 & 3.081    & 26.36 $\pm$ 0.32 &--21.58 &5.29    & 65.05            &--0.61& 0.26&70.68&0.17 & 0.69 $\pm$ 0.20& 0.57 $\pm$ 0.17 \\
 LAE3 & 3.103    & 26.73 $\pm$ 0.31 &--21.21 &7.27    &114.48   &--1.40 &0.26&46.78&0.17 & 0.66  $\pm$ 0.21& 0.59 $\pm$ 0.18 \\
 LAE4 & 3.120    & 25.82 $\pm$ 0.16 &--21.70 &3.60    &39.06     &--0.49 &0.26&106.94&0.20 &  0.82 $\pm$  0.10& 0.84 $\pm$ 0.15 \\
 LAE5 & 3.065    & 26.51 $\pm$ 0.30 &--21.85 &7.25    &66.62     &--1.07 &0.26&64.95&0.15 &  0.74 $\pm$ 0.16& 0.39 $\pm$ 0.11 \\
 stack\_all & $\ldots$ & $<28.3$ &--21.57 & $\ldots$ & $\ldots$ & $\ldots$ &0.19&40.54& 0.15 &  $<0.16$  & $<0.11$ \\
 stack\_bright & $\ldots$  & $<27.4$ &--22.67 & $\ldots$ & $\ldots$ & $\ldots$ &0.22& 89.91 & 0.12&  $<0.08$  & $<0.09$ \\
 stack\_mid & $\ldots$ &  $<28.1$ &--21.62 & $\ldots$ & $\ldots$ & $\ldots$ &0.19&44.89&0.15 &  $<0.18$  &  $<0.12$ \\
 stack\_faint& $\ldots$ &  $<27.7$ &--21.03 & $\ldots$ & $\ldots$ & $\ldots$ &0.20&22.47&0.15  &$<0.22$  &  $<0.31$ \\
 \enddata
\tablecomments{The redshifts are Ly$\alpha$ redshifts from \citet{guo2020spectroscopic}. $L_{900}/L_{1400}$ and $E(B-V)$ are obtained from our SED modeling. $f_{\rm abs,m}$ is the escape fraction given by CIGALE and $f_{\rm abs,c}$ is given by our calculation based on the Equation 2. LAE1 has a large and bright neighbor in the $Rc,i',z'$ bands, so we do not do SED modeling using its broadband photometry. The upper limits in the table are $2\sigma$ upper limits.}
\end{deluxetable*}

As we mentioned earlier, our LAE sample consists of 266 spectroscopically confirmed LAEs in the SXDS field. The $U_{\rm J}$-band image covers 246 of them. The LAEs were primarily detected and selected based on their narrowband images. They represent the most luminous galaxies in terms of Ly$\alpha$ emission, but they can be very faint in the UV continuum images. We remove 78 LAEs that are fainter than 26.2 mag in the $V$ band (see the upper panel in Figure \ref{fig:magz}). This selection criterion is determined by the detection limit of the $U_{\rm J}$-band image that sets a limit for $f_{\rm esc}$, as shown in the lower panel of Figure \ref{fig:magz}. In other words, these removed galaxies will not be detected in $U_{\rm J}$ even if their $f_{\rm esc}$ is close to 100\%. The $f_{\rm esc}$ limit is calculated using a $3\sigma$ detection limit, an average IGM absorption, and an average dust attenuation of $E(B-V)=0.2$. The detailed calculation of $f_{\rm esc}$ is presented in section 3.3. In addition, we remove another 18 LAEs that are apparently contaminated by neighboring objects. Finally, we obtain a sample of 150 LAEs and our following analyses will use this sample.
We visually inspect individual LAEs and find that some of them are slightly contaminated by their neighboring objects.
 For a given LAE, if light of its nearby objects (located outside of the photometric radius) likely reaches our photometric aperture in the $U_{\rm J}$, we simply model the objects and subtract them before we do photometry for the LAE. Deep $BVRi'$-band images are used to ensure that these nearby objects are real objects. The spatial positions of the LyC emission and UV/optical emission from galaxies are usually consistent with each other. If there are any positional offsets between the two, these offsets are typically smaller than $1\arcsec$  \citep[e.g.,][]{micheva2016searching,mostardi2013narrowband}. We use {\tt galfit} \citep{peng2002detailed} to model nearby objects, and obtain the best-fit positions and profiles simultaneously. A single S\'{e}rsic model works well here.

\subsection{Detections of Individual LyC Galaxies} 

 \begin{figure}[tb]
\centering
\includegraphics[width=0.48\textwidth]{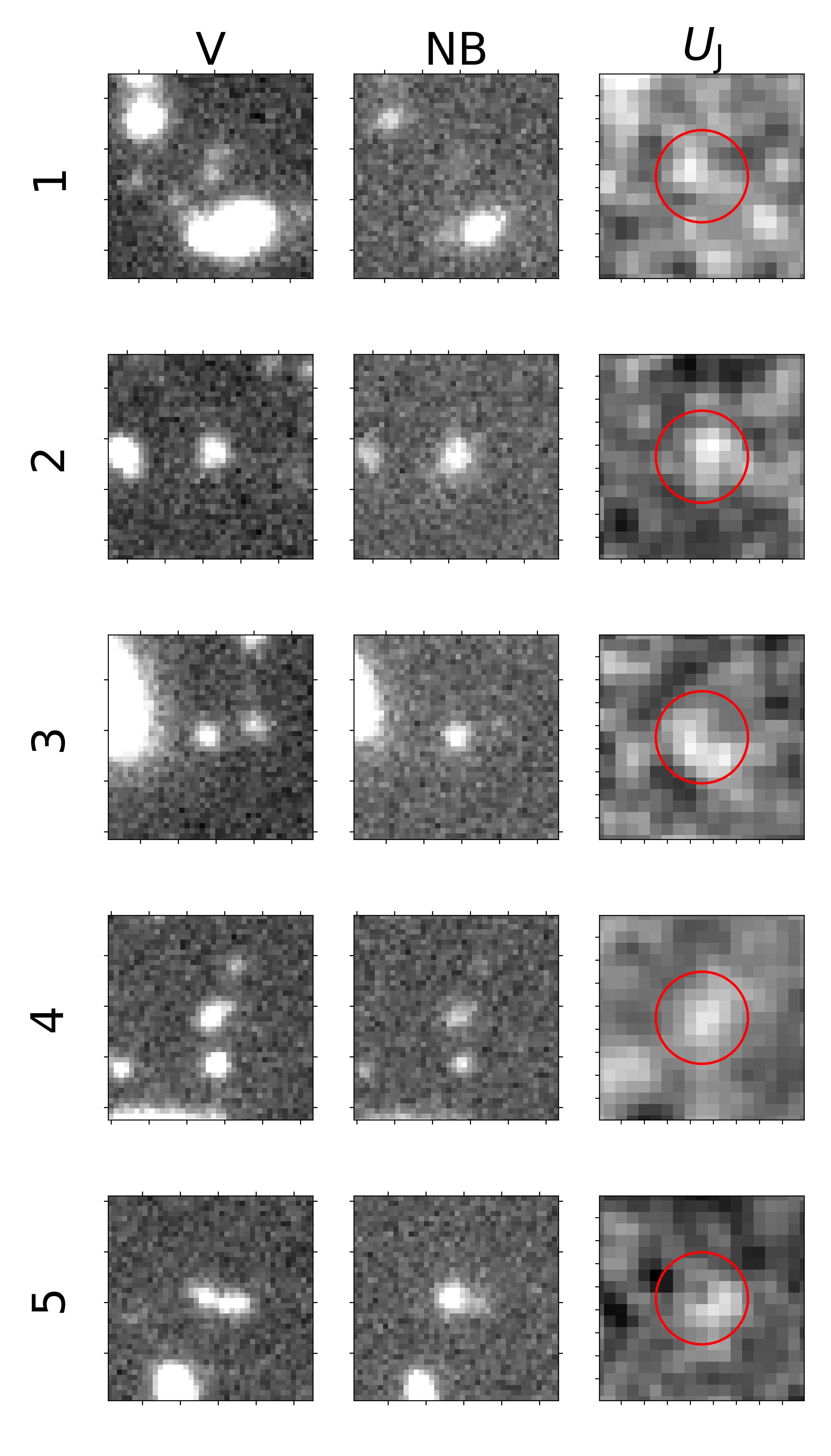}
\caption{Five LyC galaxies in the $V$, narrowband (NB497 or NB503), and $U_{\rm J}$ bands. Nearby objects in the $U_{\rm J}$ images have been modeled and subtracted. The size of the images is $8\arcsec \times8 \arcsec$. The red circles in the third column are the aperture used for photometry.}
\label{fig:3band}
\end{figure}

With the photometric measurements in the $U_{\rm J}$ band, we first identify individual LyC galaxies or candidates. As mentioned earlier, the escape fractions of LyC photons strongly depend on the IGM absorption and dust attenuation. The IGM absorption from $z\approx3$ is substantial. We adopt a criterion of $>3\sigma$ detection in $U_{\rm J}$ as the detection of a LyC candidate. With this criterion, we find 5 LyC galaxies. Figure \ref{fig:3band} shows the galaxies in three bands, $V$, narrowband (NB497 or NB503), and $U_{\rm J}$. Table 1 lists more detailed information about the LyC galaxies. The $f_{\rm esc}$ values will be derived in the following section. As shown in  Figure \ref{fig:3band} and Table 1, these LyC galaxies are very faint in $U_{\rm J}$. We also notice that the profile of individual LyC galaxies is non-Sérsic and more irregular than the UV continuum profile. A few of them may have small positional offsets between $U_{\rm J}$ and the other two bands. Such offsets have been reported previously \cite[e.g.,][]{micheva2016searching}. These phenomena may reflect clumpy and asymmetric structures that create channels for LyC photons to escape from the ISM \citep{zackrisson2013spectral,micheva2016searching}. 

We estimate the possibility of foreground contamination for the LyC galaxies using the deep Subaru imaging data. As we mentioned earlier, these images reach $\sim27.5-28.0$ mag. They also have excellent image qualities with PSF as good as $\sim0\farcs5$. Therefore, nearby objects with a separation $>0\farcs5$ can be identified and removed. We calculate the possibility of foreground objects within $0\farcs5$ from the LyC galaxies. The number counts of $z\sim3$ galaxies are obtained from the ultra-deep VIMOS $U$-band images in the GOODS-S field \citep{nonino2009deep}. Our LyC galaxies have a magnitude range from 25.5 to 26.7 mag in the $U_{\rm J}$ band and the derived surface number density is 145720 $\rm deg^{-2}$. Assuming that they are randomly distributed, the calculated probability of foreground contamination for a single object is $0.9\%$. In our total sample, the likelihood of all five detections being contaminated is 1.1\%. If we calculate the contamination rate within the aperture size of $3\farcs6$ (the size for the $U_{\rm J}$-band photometry), the contamination rate increases to $6.3\%$ for single objects, which is roughly consistent with the values in \cite{10.1111/j.1365-2966.2010.16408.x,micheva2016searching}.

\begin{figure}[tb]
\centering
\includegraphics[width=0.4\textwidth]{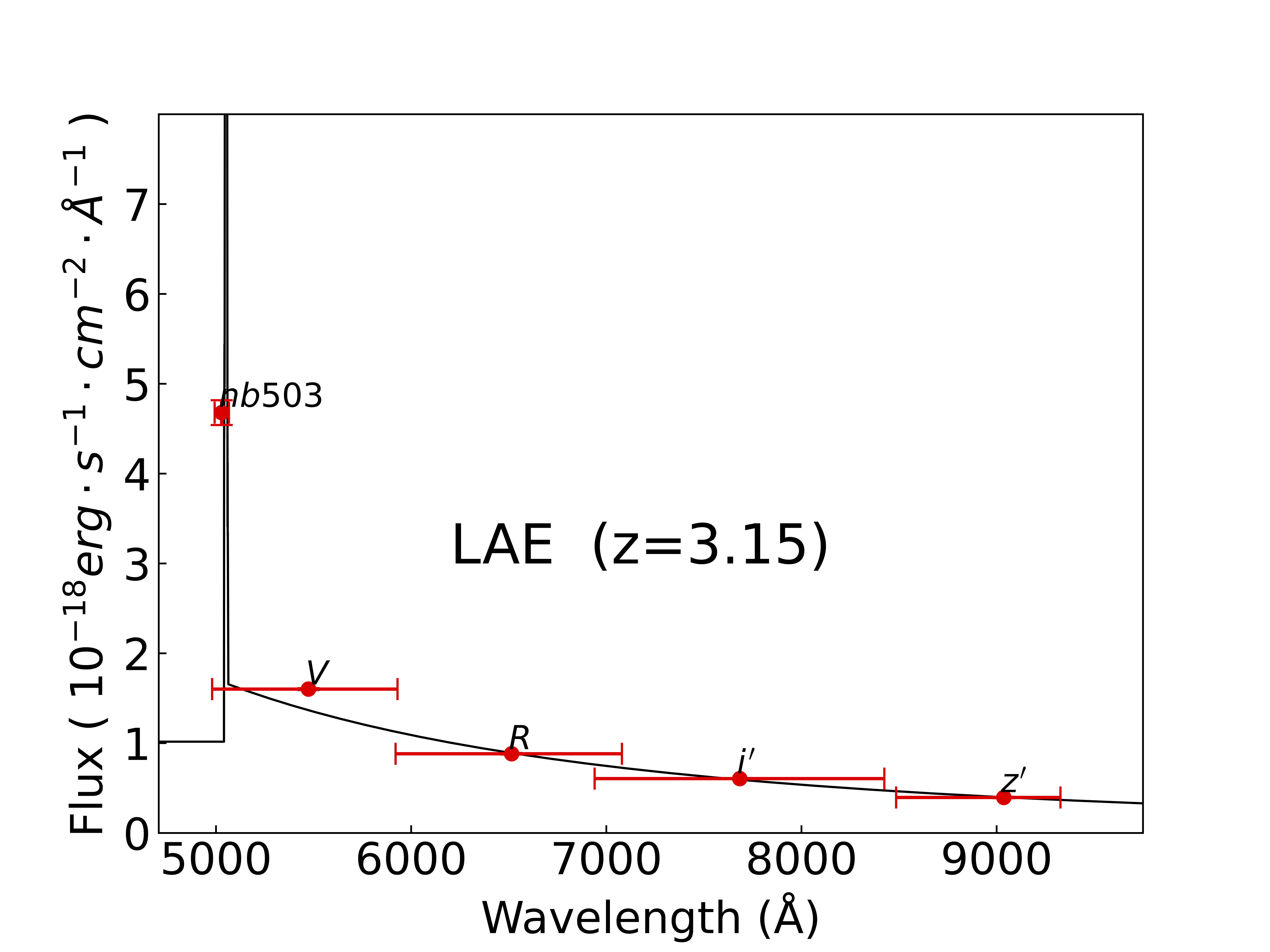}
\caption{Example of our measurement of the Ly$\alpha$ and continuum emission. The red dots show the photometric data points and the horizontal bars indicate their wavelength coverages. The solid line is the best-fit model that combines the Ly$\alpha$ and continuum emission. \label{fig:lya&conti}}
\end{figure}

We estimate non-ionizing flux at $\sim 1400$ \AA\ from the $V$-band magnitudes provided by \cite{guo2020spectroscopic}. The non-ionizing flux will be used to calculate $f_{\rm esc}$ in the next section. For $z\sim3.1$ LAEs, the Ly$\alpha$ emission line is included in the $V$ band, so we subtract the Ly$\alpha$ contribution from the $V$-band photometry and obtain non-ionizing continuum flux. We measure the Ly$\alpha$ flux using the method given in \cite{guo2020spectroscopic} and \cite{jiang2013physical}. The UV continuum and Ly$\alpha$ line emission are modeled as follows,
\begin{equation}
\label{equ:power-law}
f = \rm A \times S_{\rm Ly\alpha} +B \times \lambda^{\beta},
\end{equation}
where S$_{\rm Ly\alpha}$ is a model Ly$\alpha$ line profile,  $\beta$ is the UV continuum slope, and A and B are two scaling factors. S$_{\rm Ly\alpha}$ is obtained by co-adding a number of bright Ly$\alpha$ emission lines at $z\sim3.1$, so it has a high S/N. The UV continuum is mainly constrained by the $Ri'z'$-band photometry, and the Ly$\alpha$ line emission is mainly estimated by the narrowband photometry. In above procedure, we actually fit the five-band photometry simultaneously. More details can be found in \cite{guo2020spectroscopic}.Figure \ref{fig:lya&conti} shows an example of the flux fitting. The  results are listed in Table \ref{result}.

\subsection{Stacking Analyses} 

\begin{figure}[tb]
\centering
\includegraphics[width=0.4\textwidth]{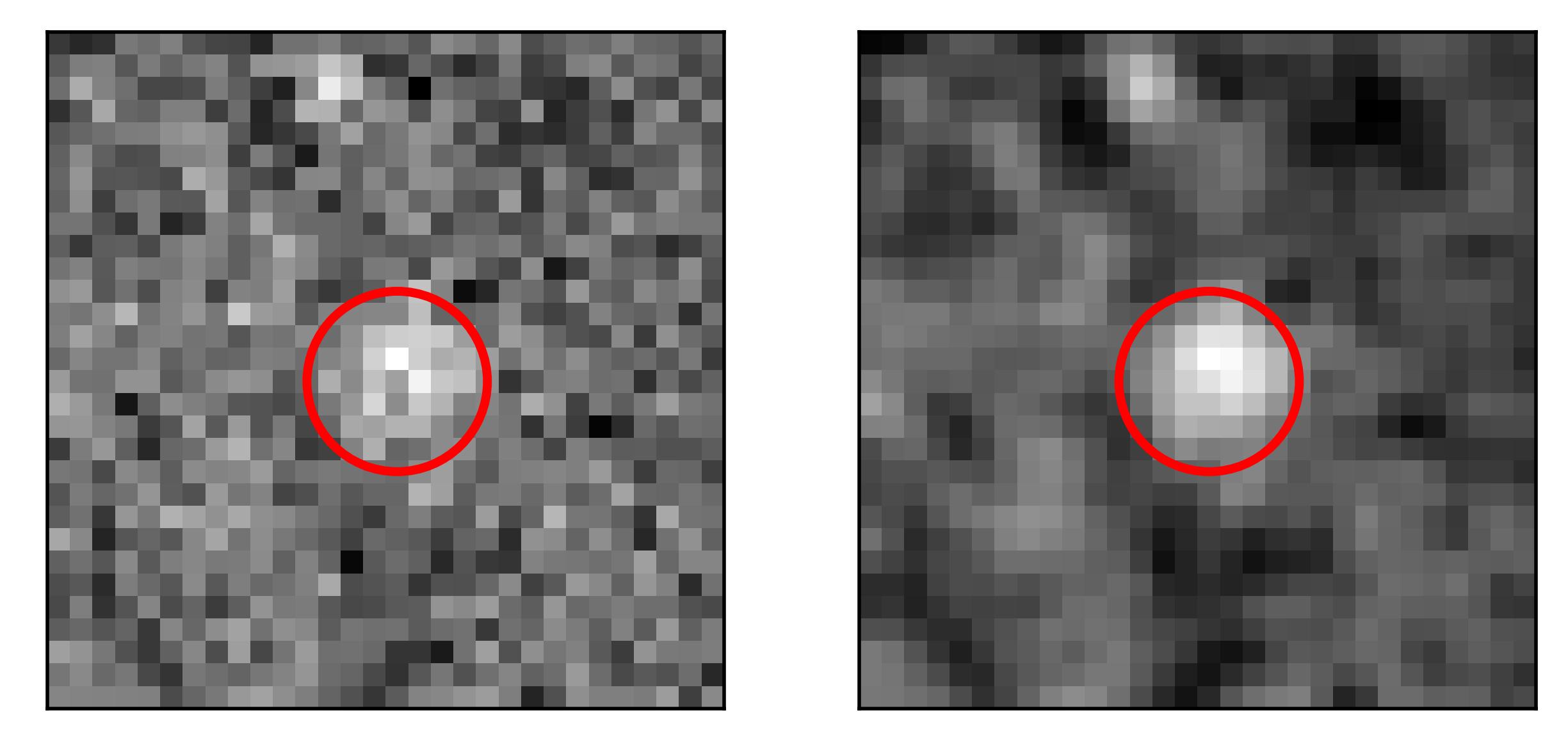}
\caption{Median stack of the  5 LyC candidates in the $U_{\rm J}$ band. The image size is $13\farcs5\times13\farcs5$. Nearby objects have been subtracted. The right panel is a smoothed image. \label{fig:stack_det}}
\end{figure}

We perform stacking analyses to enhance S/N for the LyC galaxies and to constrain $f_{\rm esc}$ for the remaining galaxies. The stacking procedure will average out the LyC emission from different galaxies and smooth out the variation of the IGM transmission along different lines-of-slights. We first stack the 5 LyC candidates, and the procedure is straightforward. We cut $U_{\rm J}$-band stamp images for the galaxies. The central positions of the stamp images are determined by the galaxy positions in the corresponding narrowband images. The stacked image is shown in Figure \ref{fig:stack_det}. The S/N of the LyC detection in this figure is greater than 8, suggesting that the individual LyC detections are reliable. 

Next, we stack the remaining 145 galaxies that are not detected in the $U_{\rm J}$ image. We first combine all 145 galaxies, and the procedure is similar to the above procedure. To efficiently remove outliers, we use sigma-clipping ($3\sigma$) to exclude the brightest and faintest pixels when combining images. No detection is found in the center of the combined image. We then divide the sample of the 145 galaxies into 3 groups, including a bright group, a medium group, and a faint group, based on their $V$-band magnitudes. The galaxies in the individual groups are combined separately, and no detection is found in any of the three combined images. The result is similar to some results in previous studies  \citep{siana2010deep,micheva2016searching,grazian2017lyman,saxena2022no}. 

For the non-detections in the stacked $U_{\rm J}$ images, we calculate their $2\sigma$ upper limits. For each stacked image, we generate $\sim3000$ mock galaxies with magnitudes ranging from 25 to 29 mag. The galaxies are randomly placed on the image. We then perform forced aperture photometry on these sources as we did for real galaxies. By fitting the measured magnitudes and uncertainties, we determine the $2\sigma$ upper limit. The results are listed in Table 1.
These limits reveal that the overall LyC escape fraction is very low. 

We also stack the $V$-band images to obtain non-ionizing flux for the three individual groups. The procedure is the same as we did for the $U_{\rm J}$-band images, and we also use sigma-clipping ($3\sigma$) to exclude outlier pixels. The measurement results (including upper limits) for ionizing and non-ionizing flux in the stacked images are 
are listed in Table \ref{result} and will be used in the following calculations.

\subsection{LyC Escape Fractions}

We calculate LyC escape fraction $f_{\rm esc}$ using the ionizing and non-ionizing flux obtained earlier. The escape fraction $f_{\rm esc}$ is defined as
\begin{equation}
\label{equ:abs}
f_{\rm esc}=\frac{f_{\rm LyC,obs}/f_{\rm UV,obs}}{f_{\rm LyC,intr}/f_{\rm UV,intr}}10^{-0.4A_{\rm UV}}e^{\tau_{\rm IGM}},
\end{equation}
where $f_{\rm LyC,obs}/f_{\rm UV,obs}$ and $f_{\rm LyC,intr}/f_{\rm UV,intr}$ are respectively the observed and intrinsic flux ratios of ionizing to non-ionizing photons, $A_{\rm UV}$ is the dust attenuation at 1400 \AA\ that can be corrected by dust attenuation law \citep{calzetti2000dust}, and $\tau_{\rm IGM}$ is the IGM optical depth. We may also define relative escape fraction $f_{\rm esc, rel}$ that assumes no dust extinction \citep{steidel2001lyman},
\begin{equation}
\label{equ:rel}
f_{\rm esc,rel}=\frac{f_{\rm LyC,obs}/f_{\rm UV,obs}}{f_{\rm LyC,intr}/f_{\rm UV,intr}}e^{\tau_{\rm IGM}}.
\end{equation}

 \begin{figure}[tb]
\centering
\includegraphics[width=0.47\textwidth]{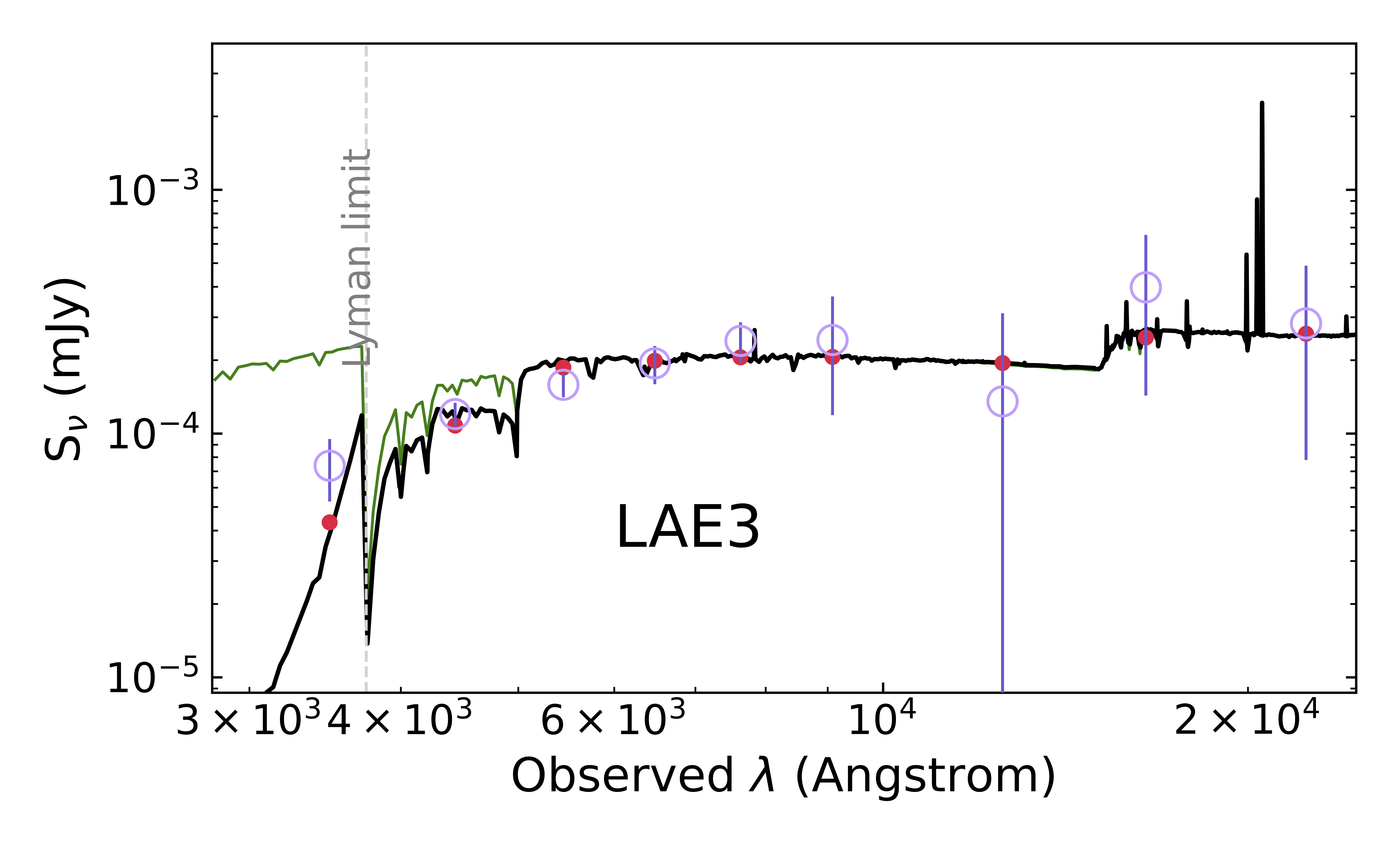}
\includegraphics[width=0.47\textwidth]{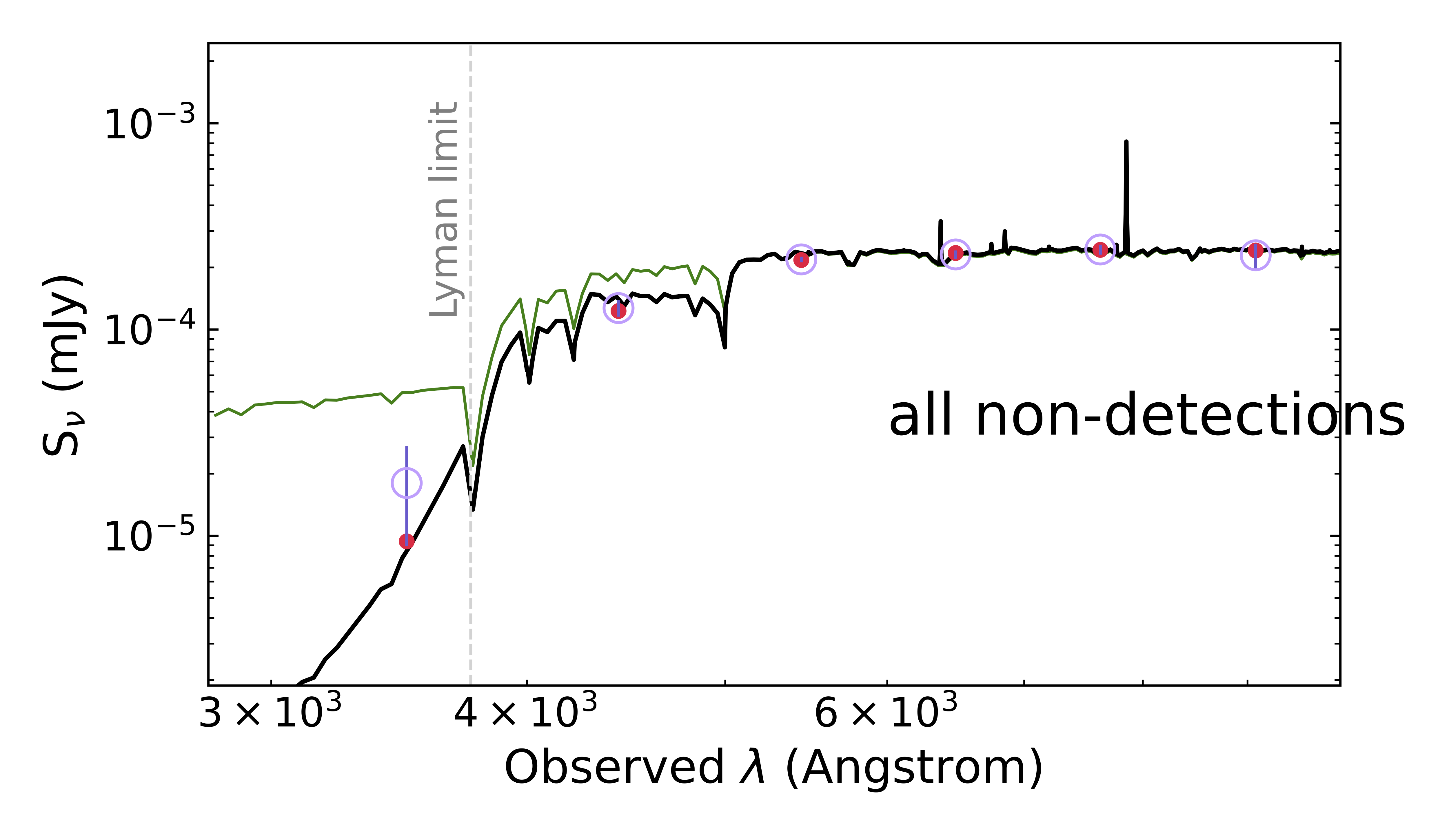}
\caption{SED modeling of LAE3 and the composite SED. In each panel, the purple circles with error bars represent observed broadband photometry, the black spectrum is the best-fit model, and the red dots are the photometry from the best-fit model. The green curve is the best-fit model without correction for the IGM absorption. We have excluded the Ly$\alpha$ emission line in the fitting process. \label{fig:sed}}
\end{figure}

Intrinsic flux ratios are roughly between 0.14 and 0.5 in star-forming galaxies with standard stellar populations and different burst ages (ranging from 1 Myr to 0.2 Gyr) \citep[e.g.,][]{bruzual2003stellar,grazian2017lyman, rivera2022bottom}. 
We perform SED modeling on individual galaxies to derive their intrinsic flux ratios and other properties. These galaxies have broadband photometry in the $B, V, R, i',$ and $z'$ bands. Their near-infrared photometric data ($J, H, K$ bands) are obtained from the UKIDSS Ultra-Deep Survey. We model their broadband SEDs using Code Investigating GALaxy Emission \citep[CIGALE;][]{boquien2019cigale,burgarella2005star}. Figure \ref{fig:sed} shows an example. We apply a SB99 model with the Salpeter IMF ($\alpha = 2.35$) and a low metallicity of $Z=0.004$. The dust reddening $E(B-V)$ is set to range from 0--1, and the \cite{calzetti2000dust} law is used for dust attenuation. We also replace the IGM absorption model in CIGALE with the model in \cite{inoue2014updated}. We mainly use the SED modeling to obtain dust attenuation and intrinsic flux ratios.  The modeling results show that the $E(B-V)$ values of our LyC galaxies are in the range of 0.1--0.25, suggesting low dust extinction in these galaxies. The derived escape fractions (denoted as $f_{\rm esc,m}$) are shown in Table \ref{result}. 

We also use Equations \ref{equ:abs} and \ref{equ:rel} to directly calculate the escape fraction ($f_{\rm esc,c}$) and apply the IGM model in \cite{inoue2014updated}. The calculated $f_{\rm esc,c}$ values are between 39\% and 84\%, comparable to the $f_{\rm esc,m}$ values of 49\%--82\% derived above. The results are shown in Table \ref{result}. As we can see, these galaxies have very high LyC escape fractions, but the fraction of such galaxies among the whole sample is only $\sim4$\%. 

Finally, we stack broadband images to produce a composite SED for all other galaxies that were not detected in $U_{\rm J}$, and then measure its $f_{\rm esc}$ by modeling this SED. The combined $U_{\rm J}$-band photometry is its $2\sigma$ upper limit. The redshifts of all the galaxies are in a small range, so the redshift different is neglected when we combine the images and model the SED. The result is $f_{\rm esc,c}<16\%\ (f_{\rm esc,m}<11\%)$. We further calculate $f_{\rm esc}$ for the three subsamples using the same method, and the values are listed in Table \ref{result}. The difference between the subsamples is primarily dependent on the sample size. These results are generally consistent with previous results in the literature, including $f_{\rm esc,rel}<12\%$ in \cite{guaita2016limits}, $f_{\rm esc}<14\%$ in \cite{micheva2016searching}, $f_{\rm esc}<9\%$ in \cite{steidel2018keck}, and $f_{\rm esc}<11\%$ in \cite{saxena2022no}. The detailed dependence of $f_{\rm esc}$ on galaxy properties will be discussed in the following section. 

It is worth noting that we have used an average or uniform IGM absorption in the above calculation of $f_{\rm esc}$. However, IGM absorption has large line-of-sight variations, and galaxies surrounded by more transparent environments tend to have higher LyC escape fractions \citep{bassett2021igm}. Therefore, the usage of a uniform IGM absorption or transmission introduces bias for individual LyC or LAE detections \citep{fletcher2019lyman, byrohl2020variations, bassett2021igm, yuan2021cdfs}. This is particularly important for the five LyC galaxies here. Following previous studies \citep[e.g.,][]{inoue2008monte, inoue2014updated}, we perform Monte Carlo simulations of the IGM transmission by generating 1000 different lines-of-sight (LoS) towards $z\sim3.1$ to estimate the uncertainty. For each LoS, we determine the column density ($N_{\rm HI}$), redshift ($z_{\rm HI}$), and Doppler parameter ($b$) distribution of the HI clouds based on the given distributions in the literature. We then calculate the average IGM transmission and dispersion as $T_{\rm IGM} = 0.23\pm0.02$ in the wavelength range covered by the $U_{\rm J}$ filter.  The results are similar to those in previous studies \citep{inoue2008monte,2016A&A...585A..48G,bassett2021igm}. The IGM absorption exhibits a stochastic nature due to its strong dependence on encounters with dense clouds ($N_{\rm HI}>10^{16}$). The final $f_{\rm esc}$ error takes into account all above errors from the photometry, SED modeling, and simulations of the IGM absorption. Our LyC candidates either have intrinsically high $f_{\rm esc}$ or have special environmental conditions. For example, a clear IGM path can be caused by a nearby AGN \citep{yuan2021cdfs}. In addition, $f_{\rm esc}$ can also be overestimated by the average model if the intrinsic ionizing to non-ionizing flux ratio is higher than usual.

\section{Discussion}

\subsection{Correlations between $f_{\rm esc}$ and Galaxy Properties}

\begin{figure*}[tb]
  \centering
  \includegraphics[width=0.85\textwidth]{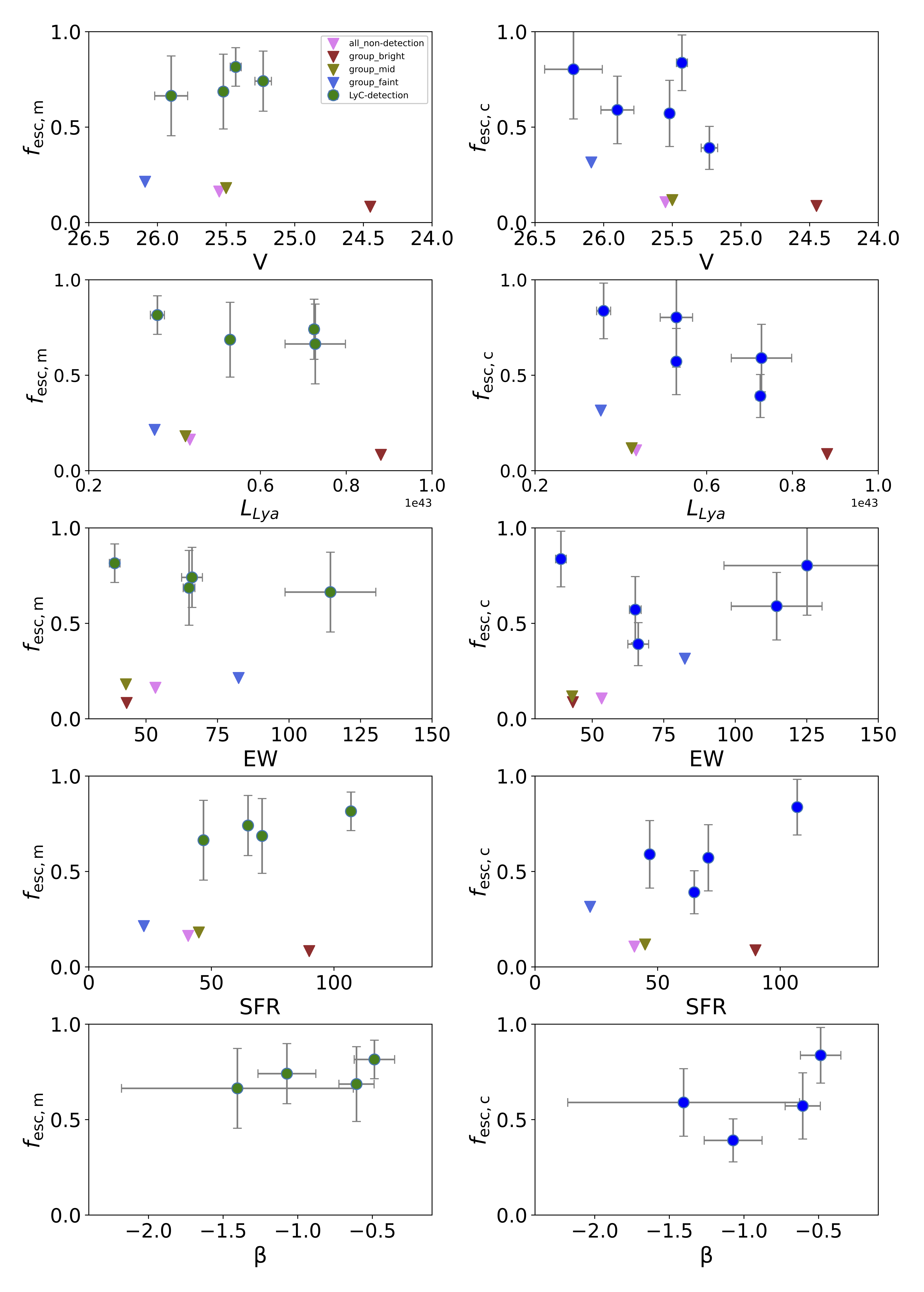}
  \caption{Relations between the escape fraction ($f_{\rm esc,m}$ and $f_{\rm esc,c}$) and other galaxy properties, including Ly$\alpha$ emission line luminosity and EW, $M_{\rm UV}$, SFR and UV slope $\beta$. The solid circles represent the five individual LyC galaxies, and the empty ones are the two possible detections. The triangles represent the upper limits from the stacked images for non-detections in $U_{\rm J}$. For Ly$\alpha$ emission and EW, the triangles show their median values (see Table \ref{result}).}
  \label{fig:abs_f_prop}
\end{figure*}

In this section we discuss possible correlations between $f_{\rm esc}$ and other galaxy properties, including UV continuum emission and continuum slope $\beta$, Ly$\alpha$ line emission and EW,  star-formation rate (SFR), etc. These properties of our galaxies are listed in Table 1. Most of them are taken from \cite{guo2020spectroscopic}, and the SFRs are from our SED modeling results. One galaxy (LAE1) is close to a much brighter object, and its photometry in long wavelengths is severely affected by this object, so it is not included in most of the following discussions. The correlation results are shown in Figure \ref{fig:abs_f_prop}. The two columns in the figure show the model-based $f_{\rm esc}$ ($f_{\rm esc,m}$) and directly measured $f_{\rm esc}$ ($f_{\rm esc,c}$), respectively. The two sets of measurements are generally consistent with each other.

We first discuss a possible selection effect for our individual LyC galaxies, as shown in the lower panel of Figure \ref{fig:magz}. Due to the flux limit in the $U_{\rm J}$ and $V$ bands, the LyC detection limit is a strong function of the $V$-band magnitude. In Figure \ref{fig:magz}, the individual LyC galaxies are all well above the limit, so this detection limit has a negligible effect on the analyses of the individual LyC galaxies. Furthermore, the limit for the stacked images is very low, so it has a negligible effect when we compare LyC galaxies and stacked non-LyC galaxies below.

Figure \ref{fig:abs_f_prop} shows that, for the individual LyC galaxies, $f_{\rm esc}$ does not significantly correlate with any other properties in the figure. This is partly due to two reasons. One is that the measurement uncertainties of the individual objects are large. The other one is these galaxies represent only a small fraction of our LAEs. \citet{steidel2018keck} and \citet{pahl2021uncontaminated} found that $f_{\rm esc}$ is positively corrected with EW. The EW range in their samples is $\sim10-40$ \AA. The LyC galaxies in our sample have much larger EWs around $70-350$ \AA, and thus we are not able to draw a conclusion based on this small sample. The correlation coefficient of the SFR and $f_{\rm esc}$ is $r\sim0.65$ while the $\beta$ is $r\sim 0.8$. The weak correlations are not confident due to large uncertainties. Therefore, we mainly compare the LyC galaxies with non-LyC galaxies shown in Figure \ref{fig:abs_f_prop} below.

Compared with the non-LyC galaxies in our sample, the LyC galaxies tend to have higher Ly$\alpha$ luminosities or EWs. This is consistent with previous results that strong Ly$\alpha$ emission is likely an indicator of the LyC leakage. We also divide the non-detections into three subsamples with the same size based on their Ly$\alpha$ EW. We still find no detection in the stacks of these subgroups and the upper limits exhibit little variations with different EW groups. Previous studies also suggest that Ly$\alpha$ luminosity or EW alone is not enough to indicate a LyC leakage \citep[e.g.,][]{bian2020lyman,ji2020hst}. The combination with a specific Ly$\alpha$ profile, such as a blue peak emission line, can be an additional proxy to search for LyC galaxies \citep[e.g.,][]{furtak2022double}. But the blue-peak feature is rare and is not found in our sample. Figure \ref{fig:abs_f_prop} also shows 
that the LyC galaxies tend to have higher SFRs. This likely reflects the higher Ly$\alpha$ luminosities or EWs above. 

Some previous studies reported a correlation between $f_{\rm esc}$ and UV slope $\beta$ in low-redshift galaxies \citep[e.g.,][]{flury2022low,chisholm2022far}. They found a strong trend that galaxies with steeper slopes have higher $f_{\rm esc}$ and suggested that LyC galaxies tend to have $\beta\le -2$. In our sample, the slopes of the LyC galaxies are roughly consistent with the non-LyC galaxies. A larger and more representative sample of LyC galaxies is needed.

\subsection{Comparison with Previous Works}

In the past decade of LyC galaxy searches, the fraction of LyC detections in star-forming galaxies is $\sim10\%$ and $f_{\rm esc}$ for the detections is about $10-40\%$ \citep{begley2022vandels}. Individual LyC galaxies usually have a wide $f_{\rm esc}$ range, such as $f_{\rm esc,rel} \sim 30\%$ of $Ion1$ (z=3.794) in \citet{ji2020hst}, $f_{\rm esc} \sim 60\%$ of $Ion3$ (z=4.0) in \citet{vanzella2018direct}, $f_{\rm esc} = 14-85\%$ in \citet{saxena2022no} $(3.11<z<3.53)$, and $f_{\rm esc} = 25-100\%$ in \citet{vanzella2012detection}. There are also upper limits of $f_{\rm esc}$ (z=3.795). For example, \cite{micheva2016searching} estimated $f_{\rm esc} < 14.7\%$ for LAEs and $<1.4\%$ for LBGs $(z\sim3.1)$. \cite{steidel2018keck} provided a large sample of spectroscopic survey of LBGs at $z\sim3$ and obtained an average value of $f_{\rm esc}\approx9\%$. Our results are in agreement with these recent works.

Previous works also suggested that the LyC escape fractions of star-forming galaxies have a bimodal distribution. The majority of star-forming galaxies have $f_{\rm esc} <10\%$, while a small fraction of them have much higher $f_{\rm esc}$. Galaxies with significant LyC detections are rare, but may provide a large fraction of ionizing photons for cosmic reionization \citep{naidu2020rapid}. Our result tends to support the claim of the $f_{\rm esc}$ bimodal distribution.

\section{Summary}

We have presented a study of LyC emission from spectroscopically confirmed galaxies at $z\sim3.1$ in the SXDS field. The galaxy sample consists of $\sim$150 LAEs and represents the brightest galaxies in terms of Ly$\alpha$ emission. 
We obtained deep UV images using a custom filter $U_{\rm J}$ that covers a rest-frame range of 810$\sim$890 \AA\ for $z=3.1$ galaxies. Five LyC galaxy candidates were detected with S/N $>3$ in $U_{\rm J}$. We estimated their LyC escape fractions using two different methods, including direct calculations and SED modeling. Their escape fractions $f_{\rm esc}$ are roughly between 40\% and 80\%. We stacked images for the remaining LAEs that were not detected in $U_{\rm J}$, but did not detect any significant signals in the stacked images. We then estimated the upper limit of their average LyC escape fraction and found $f_{\rm esc}<16$\%. Our results are generally consistent with previous studies and support a previous claim on the two populations of galaxies: most of the galaxies have small $f_{\rm esc}$ while a minor of galaxies have large $f_{\rm esc}$.

We have also measured the galaxy properties of this LAE sample, including Ly$\alpha$ emission line luminosity and EW, UV continuum luminosity and slope, and SFR. We discussed their potential correlations with $f_{\rm esc}$ and found that $f_{\rm esc}$ of individual LyC galaxies does not have a strong correlation with these properties. This is likely due to the small size of the sample and the large measurement uncertainties. Compared to the non-detections, however, the LyC galaxies have higher Ly$\alpha$ luminosities, EWs, and SFRs, which is consistent with previous studies. The correlation between UV slope and $f_{\rm esc}$ is not obvious. 

Our work adds a sample of individual LyC detections at $z\sim3$. A larger and deeper sample is needed to provide stringent constraints on the relation between $f_{\rm esc}$ and other galaxy properties. In the near future, the China Space Station Telescope \citep{zhan2021wide} will cover $\sim$400 deg$^2$ of its deep survey area in the NUV/UV and optical bands. It will provide tens of thousands of LyC detections at $z\le3$ that should be sufficient for us to achieve robust relations between $f_{\rm esc}$ and galaxy properties.

 \begin{acknowledgments}

We acknowledge support from
the National Key R\&D Program of China (2022YFF0503401),
the National Science Foundation of China (11721303, 11890693, 12225301),
and the China Manned Space Project with No. CMS-CSST-2021-A05 and  CMS-CSST-2021-A07.
Observations reported here were obtained at the MMT Observatory, a joint facility of the University of Arizona and the Smithsonian Institution.

\end{acknowledgments}

\newpage

\bibliography{ms}{}

\begin{thebibliography}{}
\expandafter\ifx\csname natexlab\endcsname\relax\def\natexlab#1{#1}\fi
\providecommand{\url}[1]{\href{#1}{#1}}
\providecommand{\dodoi}[1]{doi:~\href{http://doi.org/#1}{\nolinkurl{#1}}}
\providecommand{\doeprint}[1]{\href{http://ascl.net/#1}{\nolinkurl{http://ascl.net/#1}}}
\providecommand{\doarXiv}[1]{\href{https://arxiv.org/abs/#1}{\nolinkurl{https://arxiv.org/abs/#1}}}

\bibitem[{Aihara {et~al.}(2022)Aihara, AlSayyad, Ando, Armstrong, Bosch, Egami,
  Furusawa, Furusawa, Harasawa, Harikane, {et~al.}}]{aihara2022third}
Aihara, H., AlSayyad, Y., Ando, M., {et~al.} 2022, Publications of the
  Astronomical Society of Japan, 74, 247

\bibitem[{Allende~Prieto \& del Burgo(2016)}]{allende2016new}
Allende~Prieto, C., \& del Burgo, C. 2016, Monthly Notices of the Royal
  Astronomical Society, 455, 3864

\bibitem[{Bassett {et~al.}(2021)Bassett, Ryan-Weber, Cooke, Me{\v{s}}tri{\'c},
  Kakiichi, Prichard, \& Rafelski}]{bassett2021igm}
Bassett, R., Ryan-Weber, E., Cooke, J., {et~al.} 2021, Monthly Notices of the
  Royal Astronomical Society, 502, 108

\bibitem[{Becker {et~al.}(2015)Becker, Bolton, Madau, Pettini, Ryan-Weber, \&
  Venemans}]{becker2015evidence}
Becker, G.~D., Bolton, J.~S., Madau, P., {et~al.} 2015, Monthly Notices of the
  Royal Astronomical Society, 447, 3402

\bibitem[{Begley {et~al.}(2022)Begley, Cullen, McLure, Dunlop, Hall, Carnall,
  Hamadouche, McLeod, Amor{\'\i}n, Calabr{\`o}, {et~al.}}]{begley2022vandels}
Begley, R., Cullen, F., McLure, R., {et~al.} 2022, Monthly Notices of the Royal
  Astronomical Society, 513, 3510

\bibitem[{Bertin(2006)}]{bertin2006automatic}
Bertin, E. 2006, in Astronomical Data Analysis Software and Systems XV, Vol.
  351, 112

\bibitem[{Bertin {et~al.}(2002)Bertin, Mellier, Radovich, Missonnier, Didelon,
  \& Morin}]{bertin2002terapix}
Bertin, E., Mellier, Y., Radovich, M., {et~al.} 2002, in Astronomical Data
  Analysis Software and Systems XI, Vol. 281, 228

\bibitem[{Bian \& Fan(2020)}]{bian2020lyman}
Bian, F., \& Fan, X. 2020, Monthly Notices of the Royal Astronomical Society:
  Letters, 493, L65

\bibitem[{Boquien {et~al.}(2019)Boquien, Burgarella, Roehlly, Buat, Ciesla,
  Corre, Inoue, \& Salas}]{boquien2019cigale}
Boquien, M., Burgarella, D., Roehlly, Y., {et~al.} 2019, Astronomy \&
  Astrophysics, 622, A103

\bibitem[{Bosman {et~al.}(2022)Bosman, Davies, Becker, Keating, Davies, Zhu,
  Eilers, D’Odorico, Bian, Bischetti, {et~al.}}]{bosman2022hydrogen}
Bosman, S.~E., Davies, F.~B., Becker, G.~D., {et~al.} 2022, Monthly Notices of
  the Royal Astronomical Society, 514, 55

\bibitem[{Bouwens {et~al.}(2016)Bouwens, Smit, Labb{\'e}, Franx, Caruana,
  Oesch, Stefanon, \& Rasappu}]{bouwens2016lyman}
Bouwens, R., Smit, R., Labb{\'e}, I., {et~al.} 2016, The Astrophysical Journal,
  831, 176

\bibitem[{Bruzual \& Charlot(2003)}]{bruzual2003stellar}
Bruzual, G., \& Charlot, S. 2003, Monthly Notices of the Royal Astronomical
  Society, 344, 1000

\bibitem[{Burgarella {et~al.}(2005)Burgarella, Buat, \&
  Iglesias-Paramo}]{burgarella2005star}
Burgarella, D., Buat, V., \& Iglesias-Paramo, J. 2005, Monthly Notices of the
  Royal Astronomical Society, 360, 1413

\bibitem[{Byrohl \& Gronke(2020)}]{byrohl2020variations}
Byrohl, C., \& Gronke, M. 2020, Astronomy \& Astrophysics, 642, L16

\bibitem[{Calzetti {et~al.}(2000)Calzetti, Armus, Bohlin, Kinney, Koornneef, \&
  Storchi-Bergmann}]{calzetti2000dust}
Calzetti, D., Armus, L., Bohlin, R.~C., {et~al.} 2000, The Astrophysical
  Journal, 533, 682

\bibitem[{Chisholm {et~al.}(2022)Chisholm, Saldana-Lopez, Flury, Schaerer,
  Jaskot, Amor{\'\i}n, Atek, Finkelstein, Fleming, Ferguson,
  {et~al.}}]{chisholm2022far}
Chisholm, J., Saldana-Lopez, A., Flury, S., {et~al.} 2022, arXiv e-prints,
  arXiv

\bibitem[{{Ciardullo} {et~al.}(2012){Ciardullo}, {Gronwall}, {Wolf},
  {McCathran}, {Bond}, {Gawiser}, {Guaita}, {Feldmeier}, {Treister}, {Padilla},
  {Francke}, {Matkovi{\'c}}, {Altmann}, \& {Herrera}}]{2012ApJ...744..110C}
{Ciardullo}, R., {Gronwall}, C., {Wolf}, C., {et~al.} 2012, \apj, 744, 110,
  \dodoi{10.1088/0004-637X/744/2/110}

\bibitem[{Covey {et~al.}(2007)Covey, Ivezi{\'c}, Schlegel, Finkbeiner,
  Padmanabhan, Lupton, Ag{\"u}eros, Bochanski, Hawley, West,
  {et~al.}}]{covey2007stellar}
Covey, K., Ivezi{\'c}, {\v{Z}}., Schlegel, D., {et~al.} 2007, The Astronomical
  Journal, 134, 2398

\bibitem[{Curtis-Lake {et~al.}(2012)Curtis-Lake, McLure, Pearce, Dunlop,
  Cirasuolo, Stark, Almaini, Bradshaw, Chuter, Foucaud,
  {et~al.}}]{curtis2012remarkably}
Curtis-Lake, E., McLure, R., Pearce, H., {et~al.} 2012, Monthly Notices of the
  Royal Astronomical Society, 422, 1425

\bibitem[{De~Barros {et~al.}(2016)De~Barros, Vanzella, Amor{\'\i}n, Castellano,
  Siana, Grazian, Suh, Balestra, Vignali, Verhamme, {et~al.}}]{de2016extreme}
De~Barros, S., Vanzella, E., Amor{\'\i}n, R., {et~al.} 2016, Astronomy \&
  Astrophysics, 585, A51

\bibitem[{Dijkstra {et~al.}(2016)Dijkstra, Gronke, \&
  Venkatesan}]{dijkstra2016lyalpha}
Dijkstra, M., Gronke, M., \& Venkatesan, A. 2016, The Astrophysical Journal,
  828, 71

\bibitem[{Faisst {et~al.}(2022)Faisst, Chary, Fajardo-Acosta, Paladini,
  Rusholme, Stickley, Helou, Weaver, Brammer, Koekemoer,
  {et~al.}}]{faisst2022joint}
Faisst, A.~L., Chary, R.~R., Fajardo-Acosta, S., {et~al.} 2022, The
  Astrophysical Journal, 929, 66

\bibitem[{Fan {et~al.}(2006)Fan, Strauss, Becker, White, Gunn, Knapp, Richards,
  Schneider, Brinkmann, \& Fukugita}]{fan2006constraining}
Fan, X., Strauss, M.~A., Becker, R.~H., {et~al.} 2006, The Astronomical
  Journal, 132, 117

\bibitem[{Finkelstein {et~al.}(2012)Finkelstein, Papovich, Ryan, Pawlik,
  Dickinson, Ferguson, Finlator, Koekemoer, Giavalisco, Cooray,
  {et~al.}}]{finkelstein2012candels}
Finkelstein, S.~L., Papovich, C., Ryan, R.~E., {et~al.} 2012, The Astrophysical
  Journal, 758, 93

\bibitem[{Fletcher {et~al.}(2019)Fletcher, Tang, Robertson, Nakajima, Ellis,
  Stark, \& Inoue}]{fletcher2019lyman}
Fletcher, T.~J., Tang, M., Robertson, B.~E., {et~al.} 2019, The Astrophysical
  Journal, 878, 87

\bibitem[{Flury {et~al.}(2022)Flury, Jaskot, Ferguson, Worseck, Makan,
  Chisholm, Saldana-Lopez, Schaerer, McCandliss, Wang, {et~al.}}]{flury2022low}
Flury, S.~R., Jaskot, A.~E., Ferguson, H.~C., {et~al.} 2022, The Astrophysical
  Journal Supplement Series, 260, 1

\bibitem[{Furtak {et~al.}(2022)Furtak, Plat, Zitrin, Topping, Stark, Strait,
  Charlot, Coe, Andrade-Santos, Brada{\v{c}}, {et~al.}}]{furtak2022double}
Furtak, L.~J., Plat, A., Zitrin, A., {et~al.} 2022, arXiv preprint
  arXiv:2204.09668

\bibitem[{Furusawa {et~al.}(2008)Furusawa, Kosugi, Akiyama, Takata, Sekiguchi,
  Tanaka, Iwata, Kajisawa, Yasuda, Doi, {et~al.}}]{furusawa2008subaru}
Furusawa, H., Kosugi, G., Akiyama, M., {et~al.} 2008, The Astrophysical Journal
  Supplement Series, 176, 1

\bibitem[{Gray \& Corbally(2002)}]{gray2002spectroscopic}
Gray, R., \& Corbally, C. 2002, The Astronomical Journal, 124, 989

\bibitem[{{Grazian} {et~al.}(2016){Grazian}, {Giallongo}, {Gerbasi}, {Fiore},
  {Fontana}, {Le F{\`e}vre}, {Pentericci}, {Vanzella}, {Zamorani}, {Cassata},
  {Garilli}, {Le Brun}, {Maccagni}, {Tasca}, {Thomas}, {Zucca}, {Amor{\'\i}n},
  {Bardelli}, {Cassar{\`a}}, {Castellano}, {Cimatti}, {Cucciati}, {Durkalec},
  {Giavalisco}, {Hathi}, {Ilbert}, {Lemaux}, {Paltani}, {Ribeiro}, {Schaerer},
  {Scodeggio}, {Sommariva}, {Talia}, {Tresse}, {Vergani}, {Bonchi}, {Boutsia},
  {Capak}, {Charlot}, {Contini}, {de la Torre}, {Dunlop}, {Fotopoulou},
  {Guaita}, {Koekemoer}, {L{\'o}pez-Sanjuan}, {Mellier}, {Merlin}, {Paris},
  {Pforr}, {Pilo}, {Santini}, {Scoville}, {Taniguchi}, \&
  {Wang}}]{2016A&A...585A..48G}
{Grazian}, A., {Giallongo}, E., {Gerbasi}, R., {et~al.} 2016, \aap, 585, A48,
  \dodoi{10.1051/0004-6361/201526396}

\bibitem[{Grazian {et~al.}(2017)Grazian, Giallongo, Paris, Boutsia, Dickinson,
  Santini, Windhorst, Jansen, Cohen, Ashcraft, {et~al.}}]{grazian2017lyman}
Grazian, A., Giallongo, E., Paris, D., {et~al.} 2017, Astronomy \&
  Astrophysics, 602, A18

\bibitem[{Guaita {et~al.}(2016)Guaita, Pentericci, Grazian, Vanzella, Nonino,
  Giavalisco, Zamorani, Bongiorno, Cassata, Castellano,
  {et~al.}}]{guaita2016limits}
Guaita, L., Pentericci, L., Grazian, A., {et~al.} 2016, Astronomy \&
  Astrophysics, 587, A133

\bibitem[{Guo {et~al.}(2020)Guo, Jiang, Egami, Ning, Zheng, \&
  Ho}]{guo2020spectroscopic}
Guo, Y., Jiang, L., Egami, E., {et~al.} 2020, The Astrophysical Journal, 902,
  137

\bibitem[{Henry {et~al.}(2018)Henry, Berg, Scarlata, Verhamme, \&
  Erb}]{henry2018close}
Henry, A., Berg, D.~A., Scarlata, C., Verhamme, A., \& Erb, D. 2018, The
  Astrophysical Journal, 855, 96

\bibitem[{Hopkins {et~al.}(2007)Hopkins, Richards, \&
  Hernquist}]{hopkins2007observational}
Hopkins, P.~F., Richards, G.~T., \& Hernquist, L. 2007, The Astrophysical
  Journal, 654, 731

\bibitem[{Inoue \& Iwata(2008)}]{inoue2008monte}
Inoue, A.~K., \& Iwata, I. 2008, Monthly Notices of the Royal Astronomical
  Society, 387, 1681

\bibitem[{Inoue {et~al.}(2014)Inoue, Shimizu, Iwata, \&
  Tanaka}]{inoue2014updated}
Inoue, A.~K., Shimizu, I., Iwata, I., \& Tanaka, M. 2014, Monthly Notices of
  the Royal Astronomical Society, 442, 1805

\bibitem[{Izotov {et~al.}(2016{\natexlab{a}})Izotov, Orlitov{\'a}, Schaerer,
  Thuan, Verhamme, Guseva, \& Worseck}]{izotov2016eight}
Izotov, Y., Orlitov{\'a}, I., Schaerer, D., {et~al.} 2016{\natexlab{a}},
  Nature, 529, 178

\bibitem[{Izotov {et~al.}(2016{\natexlab{b}})Izotov, Schaerer, Thuan, Worseck,
  Guseva, Orlitov{\'a}, \& Verhamme}]{izotov2016detection}
Izotov, Y., Schaerer, D., Thuan, T., {et~al.} 2016{\natexlab{b}}, Monthly
  Notices of the Royal Astronomical Society, 461, 3683

\bibitem[{Jaskot {et~al.}(2019)Jaskot, Dowd, Oey, Scarlata, \&
  McKinney}]{jaskot2019new}
Jaskot, A.~E., Dowd, T., Oey, M., Scarlata, C., \& McKinney, J. 2019, The
  Astrophysical Journal, 885, 96

\bibitem[{Ji {et~al.}(2020)Ji, Giavalisco, Vanzella, Siana, Pentericci, Jaskot,
  Liu, Nonino, Ferguson, Castellano, {et~al.}}]{ji2020hst}
Ji, Z., Giavalisco, M., Vanzella, E., {et~al.} 2020, The Astrophysical Journal,
  888, 109

\bibitem[{Jiang {et~al.}(2015)Jiang, McGreer, Fan, Bian, Cai, Cl{\'e}ment,
  Wang, \& Fan}]{jiang2015discovery}
Jiang, L., McGreer, I.~D., Fan, X., {et~al.} 2015, The Astronomical Journal,
  149, 188

\bibitem[{Jiang {et~al.}(2013)Jiang, Egami, Mechtley, Fan, Cohen, Windhorst,
  Dav{\'e}, Finlator, Kashikawa, Ouchi, {et~al.}}]{jiang2013physical}
Jiang, L., Egami, E., Mechtley, M., {et~al.} 2013, The Astrophysical Journal,
  772, 99

\bibitem[{Jiang {et~al.}(2018)Jiang, Wu, Bian, Chiang, Ho, Shen, Zheng, Bailey,
  Blanc, Crane, {et~al.}}]{jiang2018giant}
Jiang, L., Wu, J., Bian, F., {et~al.} 2018, Nature Astronomy, 2, 962

\bibitem[{Jiang {et~al.}(2022)Jiang, Ning, Fan, Ho, Luo, Wang, Wu, Wu, Yang, \&
  Zheng}]{jiang2022definitive}
Jiang, L., Ning, Y., Fan, X., {et~al.} 2022, Nature Astronomy, 1

\bibitem[{Keating {et~al.}(2020)Keating, Kulkarni, Haehnelt, Chardin, \&
  Aubert}]{keating2020constraining}
Keating, L.~C., Kulkarni, G., Haehnelt, M.~G., Chardin, J., \& Aubert, D. 2020,
  Monthly Notices of the Royal Astronomical Society, 497, 906

\bibitem[{Konno {et~al.}(2014)Konno, Ouchi, Ono, Shimasaku, Shibuya, Furusawa,
  Nakajima, Naito, Momose, Yuma, {et~al.}}]{konno2014accelerated}
Konno, A., Ouchi, M., Ono, Y., {et~al.} 2014, The Astrophysical Journal, 797,
  16

\bibitem[{Marques-Chaves {et~al.}(2021)Marques-Chaves, Schaerer,
  Alvarez-Marquez, Colina, Dessauges-Zavadsky, Perez-Fournon, Saldana-Lopez, \&
  Verhamme}]{marques2021uv}
Marques-Chaves, R., Schaerer, D., Alvarez-Marquez, J., {et~al.} 2021, Monthly
  Notices of the Royal Astronomical Society, 507, 524

\bibitem[{Mason {et~al.}(2018)Mason, Treu, Dijkstra, Mesinger, Trenti,
  Pentericci, De~Barros, \& Vanzella}]{mason2018universe}
Mason, C.~A., Treu, T., Dijkstra, M., {et~al.} 2018, The Astrophysical Journal,
  856, 2

\bibitem[{Matthee {et~al.}(2016)Matthee, Sobral, Best, Khostovan, Oteo,
  Bouwens, \& R{\"o}ttgering}]{matthee2016production}
Matthee, J., Sobral, D., Best, P., {et~al.} 2016, Monthly Notices of the Royal
  Astronomical Society, 465, 3637

\bibitem[{Matthee {et~al.}(2015)Matthee, Sobral, Santos, R{\"o}ttgering,
  Darvish, \& Mobasher}]{matthee2015identification}
Matthee, J., Sobral, D., Santos, S., {et~al.} 2015, Monthly Notices of the
  Royal Astronomical Society, 451, 400

\bibitem[{Micheva {et~al.}(2016)Micheva, Iwata, Inoue, Matsuda, Yamada, \&
  Hayashino}]{micheva2016searching}
Micheva, G., Iwata, I., Inoue, A.~K., {et~al.} 2016, Monthly Notices of the
  Royal Astronomical Society, stw2700

\bibitem[{Mostardi {et~al.}(2013)Mostardi, Shapley, Nestor, Steidel, Reddy, \&
  Trainor}]{mostardi2013narrowband}
Mostardi, R.~E., Shapley, A.~E., Nestor, D.~B., {et~al.} 2013, The
  Astrophysical Journal, 779, 65

\bibitem[{Naidu {et~al.}(2020)Naidu, Tacchella, Mason, Bose, Oesch, \&
  Conroy}]{naidu2020rapid}
Naidu, R.~P., Tacchella, S., Mason, C.~A., {et~al.} 2020, The Astrophysical
  Journal, 892, 109

\bibitem[{{Ning} {et~al.}(2020){Ning}, {Jiang}, {Zheng}, {Wu}, {Bian}, {Egami},
  {Fan}, {Ho}, {Shen}, {Wang}, \& {Wu}}]{2020ApJ...903....4N}
{Ning}, Y., {Jiang}, L., {Zheng}, Z.-Y., {et~al.} 2020, \apj, 903, 4,
  \dodoi{10.3847/1538-4357/abb705}

\bibitem[{Nonino {et~al.}(2009)Nonino, Dickinson, Rosati, Grazian, Reddy,
  Cristiani, Giavalisco, Kuntschner, Vanzella, Daddi,
  {et~al.}}]{nonino2009deep}
Nonino, M., Dickinson, M., Rosati, P., {et~al.} 2009, The Astrophysical Journal
  Supplement Series, 183, 244

\bibitem[{Ouchi {et~al.}(2008)Ouchi, Shimasaku, Akiyama, Simpson, Saito, Ueda,
  Furusawa, Sekiguchi, Yamada, Kodama, {et~al.}}]{ouchi2008subaru}
Ouchi, M., Shimasaku, K., Akiyama, M., {et~al.} 2008, The Astrophysical Journal
  Supplement Series, 176, 301

\bibitem[{Ouchi {et~al.}(2010)Ouchi, Shimasaku, Furusawa, Saito, Yoshida,
  Akiyama, Ono, Yamada, Ota, Kashikawa, {et~al.}}]{ouchi2010statistics}
Ouchi, M., Shimasaku, K., Furusawa, H., {et~al.} 2010, The Astrophysical
  Journal, 723, 869

\bibitem[{Pahl {et~al.}(2021)Pahl, Shapley, Steidel, Chen, \&
  Reddy}]{pahl2021uncontaminated}
Pahl, A.~J., Shapley, A., Steidel, C.~C., Chen, Y., \& Reddy, N.~A. 2021,
  Monthly Notices of the Royal Astronomical Society, 505, 2447

\bibitem[{Parsa {et~al.}(2018)Parsa, Dunlop, \& McLure}]{parsa2018no}
Parsa, S., Dunlop, J.~S., \& McLure, R.~J. 2018, Monthly Notices of the Royal
  Astronomical Society, 474, 2904

\bibitem[{Peng {et~al.}(2002)Peng, Ho, Impey, \& Rix}]{peng2002detailed}
Peng, C.~Y., Ho, L.~C., Impey, C.~D., \& Rix, H.-W. 2002, The Astronomical
  Journal, 124, 266

\bibitem[{Rivera-Thorsen {et~al.}(2022)Rivera-Thorsen, Hayes, \&
  Melinder}]{rivera2022bottom}
Rivera-Thorsen, T.~E., Hayes, M., \& Melinder, J. 2022, arXiv preprint
  arXiv:2206.10799

\bibitem[{Saxena {et~al.}(2022{\natexlab{a}})Saxena, Pentericci, Ellis, Guaita,
  Calabr{\`o}, Schaerer, Vanzella, Amor{\'\i}n, Bolzonella, Castellano,
  {et~al.}}]{saxena2022no}
Saxena, A., Pentericci, L., Ellis, R., {et~al.} 2022{\natexlab{a}}, Monthly
  Notices of the Royal Astronomical Society, 511, 120

\bibitem[{Saxena {et~al.}(2022{\natexlab{b}})Saxena, Cryer, Ellis, Pentericci,
  Calabr{\`o}, Mascia, Saldana-Lopez, Schaerer, Katz, Llerena,
  {et~al.}}]{saxena2022strong}
Saxena, A., Cryer, E., Ellis, R., {et~al.} 2022{\natexlab{b}}, arXiv preprint
  arXiv:2206.06161

\bibitem[{Shapley {et~al.}(2006)Shapley, Steidel, Pettini, Adelberger, \&
  Erb}]{shapley2006direct}
Shapley, A.~E., Steidel, C.~C., Pettini, M., Adelberger, K.~L., \& Erb, D.~K.
  2006, The Astrophysical Journal, 651, 688

\bibitem[{Siana {et~al.}(2010)Siana, Teplitz, Ferguson, Brown, Giavalisco,
  Dickinson, Chary, De~Mello, Conselice, Bridge, {et~al.}}]{siana2010deep}
Siana, B., Teplitz, H.~I., Ferguson, H.~C., {et~al.} 2010, The Astrophysical
  Journal, 723, 241

\bibitem[{{Smith} {et~al.}(2018){Smith}, {Windhorst}, {Jansen}, {Cohen},
  {Jiang}, {Dijkstra}, {Koekemoer}, {Bielby}, {Inoue}, {MacKenty}, {O'Connell},
  \& {Silk}}]{smith2018hubble}
{Smith}, B.~M., {Windhorst}, R.~A., {Jansen}, R.~A., {et~al.} 2018, \apj, 853,
  191, \dodoi{10.3847/1538-4357/aaa3dc}

\bibitem[{Smith {et~al.}(2020)Smith, Windhorst, Cohen, Koekemoer, Jansen,
  White, Borthakur, Hathi, Jiang, Rutkowski, {et~al.}}]{smith2020lyman}
Smith, B.~M., Windhorst, R.~A., Cohen, S.~H., {et~al.} 2020, The Astrophysical
  Journal, 897, 41

\bibitem[{Sobral {et~al.}(2018)Sobral, Santos, Matthee, Paulino-Afonso,
  Ribeiro, Calhau, \& Khostovan}]{10.1093/mnras/sty378}
Sobral, D., Santos, S., Matthee, J., {et~al.} 2018, Monthly Notices of the
  Royal Astronomical Society, 476, 4725, \dodoi{10.1093/mnras/sty378}

\bibitem[{Steidel {et~al.}(2018)Steidel, Bogosavljevi{\'c}, Shapley, Reddy,
  Rudie, Pettini, Trainor, \& Strom}]{steidel2018keck}
Steidel, C.~C., Bogosavljevi{\'c}, M., Shapley, A.~E., {et~al.} 2018, The
  Astrophysical Journal, 869, 123

\bibitem[{Steidel {et~al.}(2001)Steidel, Pettini, \&
  Adelberger}]{steidel2001lyman}
Steidel, C.~C., Pettini, M., \& Adelberger, K.~L. 2001, The Astrophysical
  Journal, 546, 665

\bibitem[{Vacca {et~al.}(1996)Vacca, Garmany, \& Shull}]{vacca1996lyman}
Vacca, W.~D., Garmany, C.~D., \& Shull, J.~M. 1996, The Astrophysical Journal,
  460

\bibitem[{Vanzella {et~al.}(2010)Vanzella, Siana, Cristiani, \&
  Nonino}]{10.1111/j.1365-2966.2010.16408.x}
Vanzella, E., Siana, B., Cristiani, S., \& Nonino, M. 2010, Monthly Notices of
  the Royal Astronomical Society, 404, 1672,
  \dodoi{10.1111/j.1365-2966.2010.16408.x}

\bibitem[{Vanzella {et~al.}(2012)Vanzella, Guo, Giavalisco, Grazian,
  Castellano, Cristiani, Dickinson, Fontana, Nonino, Giallongo,
  {et~al.}}]{vanzella2012detection}
Vanzella, E., Guo, Y., Giavalisco, M., {et~al.} 2012, The Astrophysical
  Journal, 751, 70

\bibitem[{Vanzella {et~al.}(2016)Vanzella, De~Barros, Vasei, Alavi, Giavalisco,
  Siana, Grazian, Hasinger, Suh, Cappelluti, {et~al.}}]{vanzella2016hubble}
Vanzella, E., De~Barros, S., Vasei, K., {et~al.} 2016, The Astrophysical
  Journal, 825, 41

\bibitem[{Vanzella {et~al.}(2018)Vanzella, Nonino, Cupani, Castellano, Sani,
  Mignoli, Calura, Meneghetti, Gilli, Comastri, {et~al.}}]{vanzella2018direct}
Vanzella, E., Nonino, M., Cupani, G., {et~al.} 2018, Monthly Notices of the
  Royal Astronomical Society: Letters, 476, L15

\bibitem[{Verhamme {et~al.}(2015)Verhamme, Orlitov{\'a}, Schaerer, \&
  Hayes}]{verhamme2015using}
Verhamme, A., Orlitov{\'a}, I., Schaerer, D., \& Hayes, M. 2015, Astronomy \&
  Astrophysics, 578, A7

\bibitem[{Wang {et~al.}(2021)Wang, Heckman, Amor{\'\i}n, Borthakur, Chisholm,
  Ferguson, Flury, Giavalisco, Grazian, Hayes, {et~al.}}]{wang2021low}
Wang, B., Heckman, T.~M., Amor{\'\i}n, R., {et~al.} 2021, The Astrophysical
  Journal, 916, 3

\bibitem[{Xu {et~al.}(2022)Xu, Henry, Heckman, Chisholm, Worseck, Gronke,
  Jaskot, McCandliss, Flury, Giavalisco, {et~al.}}]{xu2022tracing}
Xu, X., Henry, A., Heckman, T., {et~al.} 2022, arXiv preprint arXiv:2205.11317

\bibitem[{Yuan {et~al.}(2021)Yuan, Zheng, Lin, Zhu, \& Rahna}]{yuan2021cdfs}
Yuan, F.-T., Zheng, Z.-Y., Lin, R., Zhu, S., \& Rahna, P. 2021, The
  Astrophysical Journal Letters, 923, L28

\bibitem[{Yung {et~al.}(2020)Yung, Somerville, Popping, \&
  Finkelstein}]{yung2020semi}
Yung, L.~A., Somerville, R.~S., Popping, G., \& Finkelstein, S.~L. 2020,
  Monthly Notices of the Royal Astronomical Society, 494, 1002

\bibitem[{Zackrisson {et~al.}(2013)Zackrisson, Inoue, \&
  Jensen}]{zackrisson2013spectral}
Zackrisson, E., Inoue, A.~K., \& Jensen, H. 2013, The Astrophysical Journal,
  777, 39

\bibitem[{Zhan(2021)}]{zhan2021wide}
Zhan, H. 2021, Chin. Sci. Bull, 66, 1290

\end{thebibliography}
\bibliographystyle{aasjournal}

\end{document}